\documentclass[review,3p,times,onecolumn]{elsarticle}
\usepackage{lineno,hyperref}
\modulolinenumbers[5]
\usepackage{comment}
\usepackage{latexsym}
\usepackage{epsfig}
\usepackage{amsmath}
\usepackage{amsfonts}
\usepackage{amssymb}
\usepackage{amsthm}
\usepackage{graphicx}
\usepackage{wrapfig}
\usepackage{pgf}
\usepackage{caption}
\usepackage{subcaption}
\usepackage{natbib}
\usepackage{tikz}
\usepackage[font=normalsize]{caption}

\graphicspath{{Figures/}}

\newcommand*{\affaddr}[1]{#1} 
\newcommand*{\affmark}[1][*]{\textsuperscript{#1}}

\journal{J.~non-Newt.~Fluid Mech.}

\begin{document}
	
	\begin{frontmatter}
			
		\medskip  
		
		\title{Fully turbulent flows of viscoplastic fluids in a rectangular duct}
		
		\medskip
		
		\author{Rodrigo~S.~Mitishita \affmark[1]} \corref{cor1} \ead{rodrigo.seiji06@gmail.com}
		\cortext[cor1]{Corresponding author} \author{Jordan~A.~MacKenzie \affmark[2]} \author{Gwynn~J.~Elfring \affmark[1]} \author{Ian.~A.~Frigaard \affmark[1,3]}
		
		\address{\affaddr{\affmark[1] Department of Mechanical Engineering, University of British Columbia, 2324 Main Mall, Vancouver, BC, V6T 1Z4, Canada} }
		
		\address{\affaddr{\affmark[2] Department of Chemical and Biological Engineering, University of British Columbia, 2360 E Mall, Vancouver, BC, V6T 1Z3, Canada} }
		
		\address{\affaddr{\affmark[3] Department of Mathematics, University of British Columbia, 1984 Mathematics Rd, Vancouver, BC, V6T 1Z2, Canada} }

\begin{abstract}
\noindent Turbulent flows of viscoplastic fluids at high Reynolds numbers have been investigated recently with direct numerical simulations (DNS) but experimental results have been limited. For this reason, we carry out an experimental study of fully turbulent flows of a yield stress fluid in a rectangular aspect ratio channel with a high-resolution laser doppler velocimetry (LDA) setup. We employ aqueous Carbopol solutions, often considered to be a simple yield stress fluid. We formulate different concentrations to address the effect of the rheology of the fluid on the turbulence statistics at an approximately constant Reynolds number. Additionally, we also perform experiments with a single Carbopol formulation at different Reynolds numbers to study its effect. The flow analysis is performed via rheology measurements, turbulence statistics and power spectral densities of velocity fluctuations. The addition of Carbopol to the flow increases turbulence anisotropy, with an enhancement of streamwise velocity fluctuations and a decrease in wall normal velocity fluctuations in comparison to water at the same mean velocity. This change is reflected on the power spectral densities of streamwise velocity fluctuations, where we observe a large increase in energy of large scale turbulent structures. Conversely, the energy of smaller scales is decreased in comparison to water, where the energy drops with a steeper scale than the Newtonian power law of $k_x^{-5/3}$. As we increase the Reynolds number with a  Carbopol solution, the streamwise Reynolds stresses approach Newtonian values in the core, which suggests diminishing effects of shear-thinning. The power spectral densities reveal that the energy content at larger scales decreases slightly with the Reynolds number. However, the shear thinning effects do not disappear even as the Reynolds number approaches 50000.

\end{abstract}


\end{frontmatter}

\section{Introduction}

Turbulent pipe flow of viscoplastic or yield stress fluids are common in a large number of industrial processes, such as transport of slurries during drilling and cementing of oil and gas wells \citep{maleki2018, bizhani2020}, particle transport applications \citep{jain2011,hormozi2017}, pulp suspension flows \citep{nikbakht2014}, and various other processing flows (e.g.~food and personal products). Although fluids used in industrial applications often exhibit more complex behaviours than simple viscoplasticity and shear-thinning, e.g.~viscoelasticity \citep{goyal2017} and time dependence/thixotropy \citep{escudier1996}, these characteristics are often neglected in flows where viscous effects are dominant \citep{guzel2009a}. By doing so, the only difference from the Newtonian behaviour is that the viscosity is a function of the strain rate \citep{bird1987}, such as inelastic, non-thixotropic yield stress liquids. These fluids are aptly named generalized Newtonian (GN) fluids, and often provide a good approximation for predicting the flow behaviour of viscoplastic fluids in practical applications. While turbulence is still an active research topic, and there are a limited number of studies with yield stress fluids.

Early studies with viscoplastic fluid flows from the 1950s have been performed via friction factor correlations. \citet{metzner1955,dodge1959} developed a methodology for predicting the friction factor of power law fluids, together with the argument that the same method should be applicable to all generalised Newtonian fluids and for different duct shapes. Application to yield stress fluids can be found in \citep{guillot1988,reed1993,founargiotakis2008}. Other authors have developed their own correlations \citep{Ryan1959,Hanks1967,Hanks1968,Wilson1985,Thomas1987}, some of which enjoy popularity in different process industries. \citet{anbarlooei2017} and \citet{anbarlooei2018} have developed friction factor correlations via a phenomenological approach based on Richardson's energy cascade and Kolmogorov's second similarity hypothesis for fully turbulent Newtonian flows, which states that in the inertial range of intermediate length scales of the energy cascade, the energy spectra of velocity fluctuations scales with a power law of -5/3 in wavenumber space, while being uniquely determined by the energy dissipation rate and independent of viscosity \citep{pope2001}. These arguments were combined with the friction factor method of \citet{gioia2006}. To formulate their theory, the authors assume that Kolmogorov's second similarity hypothesis also holds for a viscoplastic fluid. Their studies showed good agreement with Dodge \& Metzner expressions for the turbulent friction factor. However, there are discrepancies between predictions and experiments, even when the correlations have been experimentally validated \citep{dodge1959,escudier1996,presti2000,peixinho2005}. Many of the phenomenological theories also are used to predict turbulent transition, again with noted differences \citep{Nouar2001,guzel2009b}. 

The discrepancies between experiments with viscoplastic fluids and friction factor correlations may come from limited rheological measurements in early years and from the materials used experimentally, which include slurries \citep{kelessidis2011} and pulp suspensions \citep{nikbakht2014}. Viscoelasticity or thixotropy effects are not usually accounted for, but to some extent are always present in non-Newtonian flows. For example, when flexible polymers are added to water, the resulting viscoelasticity can significantly reduce frictional drag in turbulent flow, often near 70 \% \citep{warholic1999influence, white2008, escudier2009, jaafar2009, graham2014}. It is perhaps only in recent decades that researchers have settled on specific fluids that are better suited as a relatively simple yield stress fluid. By far the most widely used polymer additive is Carbopol \citep{dinkgreve2018}. When correctly mixed, it provides a clear solution which is ideal for visualization and laser velocimetry experiments \citep{jossic2013, peixinho2005, putz2008, guzel2009a}, while exhibiting little viscoelastic behaviour beyond the yielding point and negligible thixotropy \citep{jossic2013, dinkgreve2018, daneshi2020}, with only the yield stress and shear-thinning as dominant characteristics. Thus, Carbopol solutions have been considered to be very close to an ideal viscoplastic fluid and are also used in our study.

Non-intrusive flow velocimetry techniques such as particle image velocimetry (PIV), laser doppler anemometry (LDA) and ultrasound doppler velocimetry (UDV) advanced the field significantly. These techniques have been widely used over the years in turbulence measurements to provide high-resolution measurements of velocity fluctuations, combined with pressure and flow rate measurements and rheological characterization of the working fluids. Notably, velocimetry experiments provided valuable information on the transition to turbulence, which is still an active research topic with both Newtonian and non-Newtonian fluids. \citet{peixinho2005} observed that turbulent puffs develop more frequently with an increase in Reynolds number, eventually becoming turbulent once the puff frequency is high enough, i.e.~a saturation phenomenon. \citet{guzel2009a} concluded that fully turbulent flows in viscoplastic fluids are achieved only when the yield stress is exceeded by the fluctuating Reynolds stresses, across the entirety of the pipe. Using data from \citep{guzel2009a}, \citet{guzel2009b} concluded that an estimation of transitional Reynolds numbers can be predicted from a radial averaged Reynolds number from the time-averaged velocity profile. The average turbulence intensity across the pipe must be sufficiently high to overcome the yield stress. An interesting rheological effect is the asymmetry of the average velocity profile during transition, as seen in experiments with viscoplastic fluids \citep{peixinho2005, escudier1996,guzel2009a} and in viscoelastic drag-reducing polymer solutions \citep{jaafar2010, guzel2009a}.  

LDA experiments by \citet{peixinho2005} showed results of turbulent pipe flow in Carbopol solutions and found an upturn in the velocity profile scaled in inner units, and an increase in the streamwise Reynolds stresses in comparison to water, similar to drag-reducing solutions. The Carbopol used by \citet{peixinho2005} showed strong drag reduction behaviour, more than a CMC solution also used in the study. Whether the drag reduction in their Carbopol is evidence of significant elasticity or shear-thinning is not clear \citep{pinho1990,rudman2004}. \citet{presti2000} also observed streamwise turbulent intensities being larger than the Newtonian counterpart (similar to \citet{peixinho2005}) near the wall and radial turbulent intensities being similar to the Newtonian case. As the Reynolds number increases, streamwise and wall normal velocity fluctuations also seem approach the Newtonian values, likely due to the increased viscous response to high frequency turbulence. A similar Reynolds number dependence was also observed during the turbulent flow of a Laponite solution, a thixotropic, viscoplastic fluid in the experiments by \citet{escudier1996}. More details about the high frequency dynamics can be studied through the frequency spectra of velocity fluctuations of viscoplastic fluids, which has not been investigated experimentally, to the best of our knowledge. 

With all the above complexities, it is interesting to reflect that the most accurate way to study the fully turbulent flow of a GN fluid is via direct numerical simulation (DNS). These studies have become more common in the past decade but still lag behind those performed for Newtonian and viscoelastic fluids. Resolution of models such as the Bingham or Herschel-Bulkley fluid models is prohibitively expensive for DNS because of the singular viscosity at zero shear rate \citep{rudman2004}. Instead most of high Reynolds number DNS results that we know have been computed using regularized versions of these models. Friction factor results from the DNS of a power-law fluid in \citep{singh2018} agree well with the measurements of \citet{dodge1959}. DNS can also provide turbulence measurements that are challenging to obtain experimentally such as the energy budgets, which require data on all three components of velocity, pressure fluctuations and also viscosity fluctuations in non-Newtonian fluids specifically. In the work of \citet{singh2017b}, the turbulent kinetic energy budgets are presented for pipe flow of power-law fluids, where the amount of shear-thinning is quantified by the decrease in the power-law index $n$, with with $0 < n < 1$. The dependence of Reynolds stresses on the power law index $n$ is found mainly near the wall, where the velocity gradients and also fluctuations of shear rate are larger. Additionally, the authors observed a decrease in the turbulent production term, and an increase of the energy dissipation term as $n$ decreases. Similarly, \citet{singh2018} analyze Reynolds number effects in turbulent flows of shear thinning fluids. They conclude that the effects of Reynolds number on velocity fluctuations are more apparent only in the buffer layer and viscous sublayer, with increased turbulence production and viscous dissipation as $Re$ increases.

Shear-thinning effects can play a large role in turbulence dynamics, as shown by DNS studies \citep{rudman2004,singh2017a,singh2017b}. The decrease in power-law index in a Herschel-Bulkley fluid can increase the root mean square of velocity fluctuations $u_{rms}$ and give a small decrease in Reynolds stress. These changes in turbulent intensities are a consequence of the enhancement of more elongated turbulent structures and the attenuation of smaller scale vortices \citep{singh2017b}. a similar effect to that observed in turbulent drag reduction with viscoelastic fluids \citep{white2008}. The investigation by \citet{rosti2018} is notable for including viscoelasticity in the laminar regime via a Kelvin-Voigt modification of the Bingham constitutive equation without use of regularization, but the turbulent regime had negligible viscoelasticity with a relatively low Reynolds number. Their results highlight that the yield stress can effectively dampen turbulence structures, with larger unyielded volumes of fluid near the core as the yield stress increases. As the yield stress becomes larger, streamwise turbulent structures are also enhanced, and small scale structures are suppressed \citep{singh2017a, rosti2018}.

While these DNS studies are breaking new ground, the experimental results available are limited and often have used fluids with a more complex rheology than Carbopol, e.g.~dilute viscoelastic fluids. A quantitative analysis of the effect of Reynolds number and the most dominant rheological features of viscoplastic fluids (the yield stress and shear-thinning viscosity) has not been performed in detail with experiments. This is likely due to the difficulties of achieving a turbulent flow by pumping a viscoplastic fluid in experiments. To account for this gap, we present an experimental investigation of turbulent flows with Carbopol solutions at three different concentrations. We present velocity and Reynolds stresses measured with a two component LDA setup, with high spatial and temporal resolution adequate to resolve the small scales in the inertial range. We also provide a comparison of spectral analysis between water and Carbopol solutions, to assess the changes in energy content of different eddy length scales. Such an analysis has not yet been performed with yield stress fluids experimentally or via DNS, to the best of our knowledge. In the turbulent regime, Carbopol solutions are often considered inelastic, and thus the viscous or shear-thinning contribution to the flow is considered dominant \citep{guzel2009a}. Therefore, we focus our analysis on the effect of the shear thinning behaviour, in addition to the possible influence of the yield stress. 

The paper is organized as follows. In \hyperref[Exp]{Section~\ref*{Exp}} we outline the experimental methods and procedures with both LDA and also rheometry. The results are organized in two sections: In \hyperref[Results1]{Section~\ref*{Results1}}, the effect of the Carbopol concentration in the turbulent flow is investigated via rheological characterization of the test fluids, statistical methods and spectral analysis at a similar Reynolds number. Then, \hyperref[Results2]{Section~\ref*{Results2}} presents an analysis of the effect of the Reynolds number in the flow statistics and spectra, while keeping the Carbopol concentration the same. The major contributions of the paper are summarized in \hyperref[Conclusion]{Section~\ref*{Conclusion}}.


\section{Experiments} \label{Exp}

\subsection{Flow loop}

\begin{figure}
	\centering
	\includegraphics[width=\textwidth]{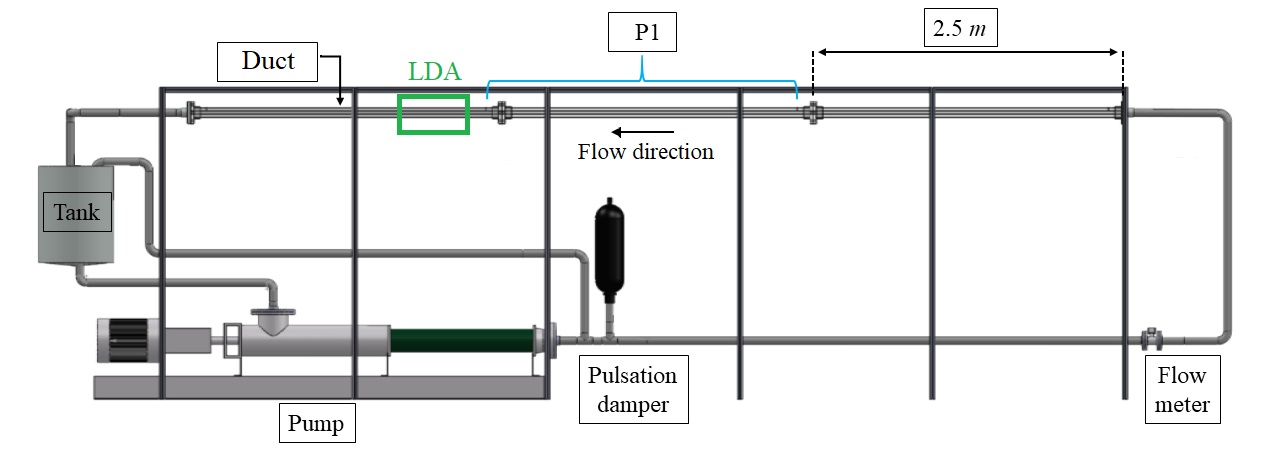}
	\caption{\fontsize{9}{9}\selectfont Flow loop schematic.}
	\label{flowloop}
\end{figure}

We carry out our turbulent flow investigation in a horizontal, pump driven flow loop through a transparent $7.5 \, m$ test section with a rectangular cross section, connected to the pump discharge pipeline by directional valves. A schematic of the flow loop is presented in \hyperref[flowloop]{Figure~\ref*{flowloop}}. The rectangular test section (``Channel'' in \hyperref[flowloop]{Figure~\ref*{flowloop}}) is made of three 2.5 $m$ long, clear acrylic channels connected together by flanges. The inner dimensions of the channel are width $W = 50.8$ \textit{mm} x height $H = 25.4$ \textit{mm} $= 2h$, where $h$ is the channel half-height, with hydraulic diameter $D_h = 2WH/(W+H)$ of 33.8 $mm$. The flow in the storage tank is driven by a Netzsch NEMO progressing cavity pump, capable of a maximum flow rate of 20 $l/s$. This type of pump minimizes shearing of the fluid, thus decreasing the degradation rate of polymer solutions. A bypass pipe connects the the pump discharge to the tank. A Parker pulsation damper was installed after the pump discharge to prevent pressure fluctuations. An Omega FMG 606 magnetic flow meter, with accuracy of 0.2 \% of full scale is used to measure the average flow rate, and the average velocity can be calculated by dividing the flow rate by the cross sectional area of each test section. The pressure drop along 2.5 \textit{m} of the rectangular test section is measured by an Omega DPG 409 differential pressure transducer (P1 in \hyperref[flowloop]{Figure~\ref*{flowloop}}), with 0.08 \% accuracy and data acquisition rate of 3 \textit{hz}. The temperature of the fluid in the tank is measured by an Omega TC-NPT pipe thermocouple, of 0.5 \% accuracy. We do not control the fluid temperature in the loop, but the temperature increases by less than 2$^\circ$C during a full velocity profile measurement. Remote control of the pump speed, as well as the signal acquisition from the thermocouple, pressure transducers and flow meter are provided by the National Instruments LabVIEW software and compact data acquisition modules. We perform laser doppler velocimetry (LDA) experiments at over 5 \textit{m} (or approximately 150$D_h$) downstream of the test section inlet, where we consider the flow to be fully developed.

\subsection{Laser doppler anemometer setup} \label{LDA}

\begin{figure}
	\centering
	\includegraphics[width=\textwidth]{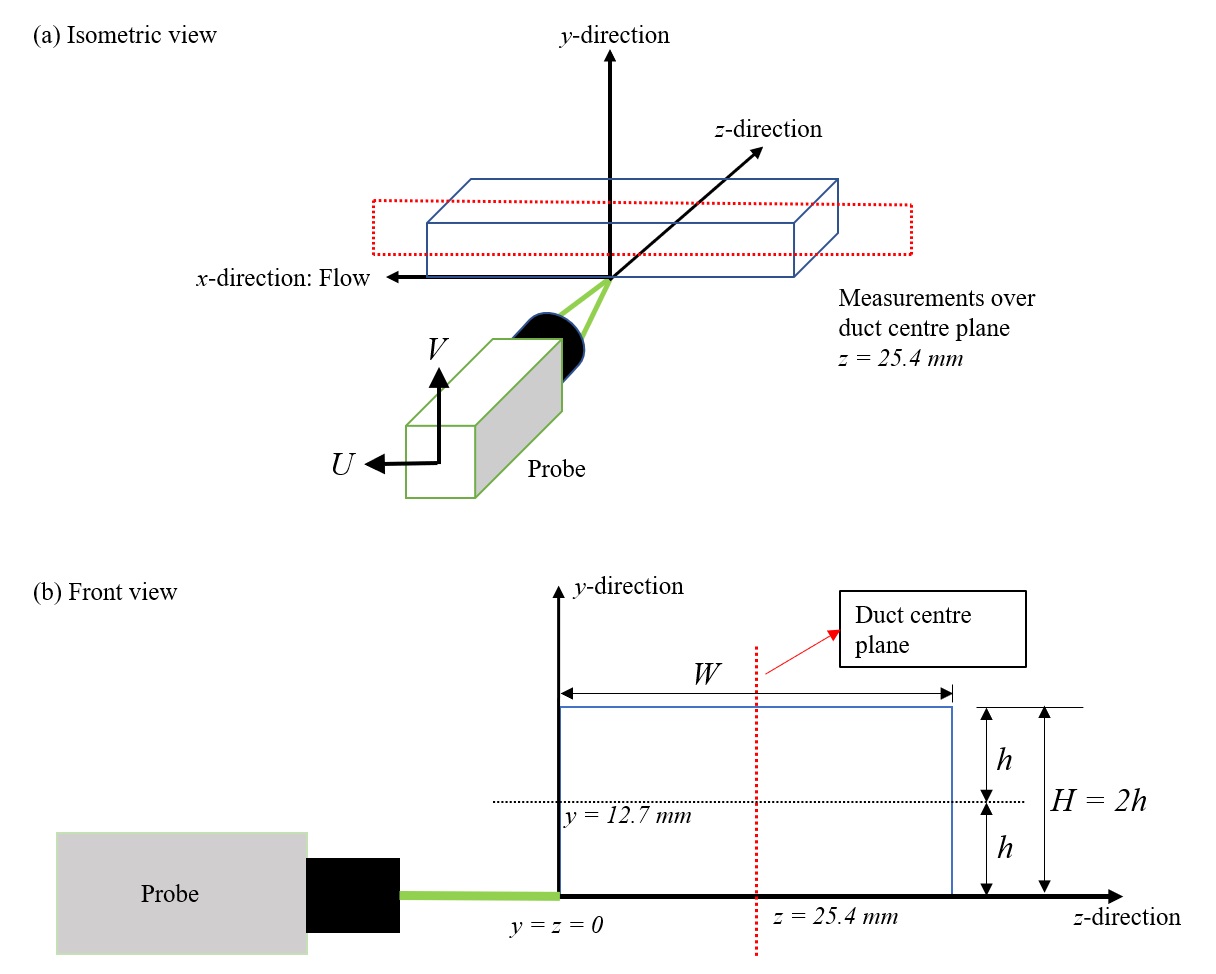}
	\caption{\fontsize{9}{9}\selectfont Three-dimensional (a) and front view (b) schematics of the LDA measurement setup in the channel. The channel centre plane, or the measurement plane, is shown in dotted lines.}
	\label{LDA_schematic}
\end{figure}

To acquire turbulent velocity data in the Carbopol solutions, we perform instantaneous, point-wise velocity measurements with a two-component Dantec Dynamics FlowExplorer LDA setup in back-scatter mode. The receiving optics and laser system are contained in a probe. The laser source supplies a pair of 532 \textit{nm} wavelength (green) laser beams, separated horizontally by a distance of 60 \textit{mm} for velocity measurements in the streamwise direction (or \textit{x}-direction), and a pair of 561 \textit{nm} wavelength (yellow) laser beams, separated vertically by a distance of 60 \textit{mm} for velocity measurements in the wall normal  direction (\textit{y}-direction). For clarity, the LDA measurement schematic is shown in \hyperref[LDA_schematic]{Figure~\ref*{LDA_schematic}} (a), along with main flow coordinates. The \textit{x}-direction corresponds to the main flow direction of positive streamwise velocity $U$. The \textit{y} coordinate represents the direction of $V$ or wall normal velocity, with the reference zero point on the bottom wall. Therefore the positive \textit{y} direction corresponds to fluid moving away from the bottom wall at $y/h = 0$ to the centre $y/h = 1$. The \textit{z} coordinate corresponds to spanwise directions with the reference zero point on the front wall. Therefore the measurement plane, which is the channel centre plane, is located at $z/W = 0.5$ or $z = 25.4$ \textit{mm}. These details illustrated in \hyperref[LDA_schematic]{Figure~\ref*{LDA_schematic}} (b). The optical setup allows for an ellipsoidal measurement volume of 0.1 \textit{mm} diameter and 0.3 \textit{mm} length with a 150 \textit{mm} focal length lens. A frequency shift of 80 \textit{Mhz} is applied to each laser by a Bragg cell. The probe is connected by to the Burst Spectrum Analyser (BSA) signal processor for data acquisition with the Dantec BSA software. Seeding particles manufactured by Dantec Dynamics, of 5 $\mu$\textit{m} average size, are used for the experiments. The uncertainty of the LDA system measurements, according to the factory calibration certificate, is 0.1 \%. 

Considering that LDA measures fluid velocity at a point, the probe requires a traverse system for velocity profile measurements. Our custom traverse system consists of a linear motorized actuator by Zaber Motion Control of a total travel distance of 100 \textit{mm} with 25 $\mu$\textit{m} accuracy for vertical motion, to allow traverse in the \textit{y}-direction. Attached to the vertical traverse is another linear actuator by Zaber for horizontal movement in the \textit{z}-direction in \hyperref[LDA_schematic]{Figure~\ref*{LDA_schematic}}. This actuator has a total travel distance of 250 \textit{mm} with 63 $\mu$\textit{m} accuracy. We estimate the position of the walls ($y = 0$ and $z = 0$ in \hyperref[LDA_schematic]{Figure~\ref*{LDA_schematic}}) with the LDA probe by monitoring the photomultiplier anode current value while traversing from a position within the channel towards the wall. We record the zero point where the anode current reading shows very high values due to reflection of the laser beams, as the measurement volume approaches the wall location. The accuracy of the bottom (via \textit{y}-direction traverse) and front wall (via \textit{z}-direction traverse) location measurement is assumed to be 0.1 \textit{mm} and 0.3 \textit{mm}, which is the diameter and length of the measurement volume respectively. The effect of refraction of the laser beams while traversing in the \textit{z}-direction is calculated with the refractive indices of both Carbopol or water ($n_r = 1.33$) and the clear plexiglass ($n_r =  1.49$). This results in a measurement volume traverse of 0.691 \textit{mm} within the channel as the probe moves 0.5 \textit{mm}. 

\subsection{Rheometry} \label{kinexus}

Rheological measurements are performed to characterize Carbopol solutions prepared for the flow loop experiments. A high-resolution Malvern Kinexus Ultra+ rheometer with a set of roughened parallel plates with 40 \textit{mm} diameter is employed for all rheology experiments to prevent wall slip, which is common in yield stress fluids. The samples used in the rheometer were taken from the flow loop tank after a full traverse with the LDA at a constant bulk velocity. The rheometry is performed at the same temperature as the average recorded from the flow loop during the experiments. Prior to every rheolometry test, we perform a pre-shear at $\dot{\gamma} = 100 s^{-1}$ for 30 \textit{s} followed by a 30 \textit{s} controlled stress rest at 0 \textit{Pa}. We perform shear-rate controlled ramp-up/down tests with a gap height of 1 \textit{mm} to assess thixotropy and measure the viscosity of the Carbopol solutions at low shear rates. Additionally, these experiments are useful to estimate the dynamic yield stress of the fluids via a fit to the Herschel Bulkley model for viscoplastic fluids:

\begin{equation} \label{HB}
	\tau = \tau_y + K\dot{\gamma}^n,	
\end{equation}

\noindent where $\tau$ is the shear stress [Pa], $\tau_y$ is the yield stress [Pa], $K$ is the consistency index [Pa s$^n$] and $n$ is the power law index [-]. Note that the \hyperref[HB]{Equation~\ref*{HB}} is valid only if $\tau > \tau_y$, because if $\tau < \tau_y$ then $\dot{\gamma} = 0$. To measure the viscosity at high shear rates, a steady state flow curve test is performed with a gap of 0.3 \textit{mm} to minimize inertial effects. Next, stress-controlled amplitude sweeps at 0.5 \textit{hz} were carried out to measure the values of the storage ($G'$ [Pa]) and loss ($G''$ [Pa]) moduli. Finally, we perform frequency sweeps from 1 to 0.01 \textit{Hz} in the linear viscoelastic regime.

\subsection{Carbopol mixing and flow loop experiment procedure} \label{procedure}

Before showing the main turbulence results, we present the procedure for mixing the Carbopol solutions, followed by the flow loop data acquisition procedure with the LDA, pressure, temperature and flow rate sensors. The mixing procedure is similar to \citet{bizhani2020}. The preparation of the Carbopol solution consists in mixing concentrated batches of 30 \textit{l} of water at a specific concentration of the Lubrizol Carbopol EZ-2 powder. The concentrated solutions were mixed in a bucket with a three blade impeller at 150 rpm for a whole day. Each specific concentration of Carbopol is measured relative to the total desired volume of 220 \textit{l} of water. Next, the concentrated solution at a given concentration and 190 \textit{l} of water are added to the flow loop tank. This solution is acidic and needs to be neutralized with NaOH to build viscosity and yield stress. With the fluid mixing at a constant pump speed, an aqueous solution of sodium hydroxide of 1/3.5 NaOH to Carbopol weight is added to the tank, and mixed for 45 minutes for homogenization. Three concentrations of Carbopol were used in our experiments: 0.06\%, 0.08\% and 0.10\%.

The main LDA velocity measurements were performed at bulk velocity $U_b = 3.8$ \textit{m/s} for 0.06\% Carbopol, $U_b = 5.8$ \textit{m/s} for 0.08\% and $U_b = 7.0$ \textit{m/s} for 0.10\%, with $U_b$ calculated from the flow meter measurements. As we show in the next section, these velocity parameters allow us to investigate the effect of changes in rheology in the turbulent flows at somewhat similar Reynolds numbers of the order of 20000. Higher velocities were needed for the more concentrated solutions to achieve a fully turbulent flow. Additional experiments with the 0.06\% Carbopol were performed at $U_b = 2.8$ and $5.8$ \textit{m/s}, to analyze the influence of the Reynolds number with a constant Carbopol concentration. For comparison, we also performed LDA experiments with water at bulk velocities $U_b = 2.8, 3.8$ and $5.8$ \textit{m/s} and only pressure, flow rate and temperature measurements at $U_b = 7.0$ \textit{m/s}.

We consider the velocity profile to be symmetric along the centre plane of the channel during fully turbulent flow, since reports of asymmetry in the velocity profile have been reported during transition to turbulence in Carbopol solutions \citep{peixinho2005,guzel2009a}. Thus, a traverse to measure half of the total velocity profile corresponds to measurements on the centre plane of the channel (\hyperref[LDA_schematic]{Figure~\ref*{LDA_schematic}}), from the bottom wall at $y/h = 0$ to the centreline. Taking this into account with the Carbopol solutions, we measure the velocity at each position in the y-direction until at least 50000 measurements are acquired for the $U$ component, and at least 30000 points for the $V$ component in non-coincidence mode. Coincidence mode velocity measurements with Carbopol were limited to 30000 data points due to the lower data rate. At least 50000 data points were measured for all water measurements. A verification of statistical convergence is reported in the supplementary material.

The temperature, flow rate and pressure drop are recorded at the same time as the LDA experiments. The pressure drop measurements $\Delta P$ are used to calculate the mean wall shear stress $\tau_w = \Delta P D_h/4 L$, where $L$ is the length between each pressure tap. With $\tau_w$ we define the friction velocity $u_{\tau} = \sqrt{ \tau_w/ \rho }$. The density of Carbopol is the same as water ($\rho$ = 999 \textit{kg/m$^3$}). With the bulk velocity and the friction velocity we define the generalized Reynolds number and the frictional Reynolds number, $Re_G = \rho U_b D_h/ \eta_w$ and  $Re_\tau = \rho u_\tau h/ \eta_w$ respectively, where $\eta_w = \tau_w/\dot{\gamma}_w$ is the viscosity of the fluid at the wall and $h$ is the channel half-height, which is also the boundary layer height in fully developed channel flow. The mean shear rate at the wall $\dot{\gamma}_w$ can be obtained with the wall shear stress and rheometer flow curve data fit to a constitutive equation. This definition of $\eta_w$ has been employed in a few other experimental works such as \citet{escudier2009} and \citet{owolabi2017} given the difficulty to establish the Reynolds number with shear-thinning fluids, where the viscosity varies across the channel. The viscosity at the wall is also used to define the wall unit $y^+_0 = \eta_w /\rho u_{\tau}$, a viscous length scaled used to normalize the wall normal coordinate $y$, measured from the bottom wall of the channel.

The scope of our experimental work is to analyze the effect of the rheological changes on the statistics of the turbulent channel flow in \hyperref[Results1]{Section~\ref*{Results1}}. We also study the effects of a relatively wide range of Reynolds numbers with the same Carbopol concentration in \hyperref[Results2]{Section~\ref*{Results2}}. We present the main flow loop parameters in \hyperref[parameters]{Table~\ref*{parameters}}, for the experiments with water and all concentrations of Carbopol solutions.

\begin{table}[!htb]
\fontsize{10}{10}\selectfont
\centering
\begin{tabular}{  c  c  c  c  c  c }
	\hline	
	U$_b$ [$m/s$] & $Re$ [-] & $u_\tau$ [$m/s$] & $Re_{\tau}$ [-] & T [$^\circ$C]  & $\tau_w$ [$Pa$] \\
	\hline
	\multicolumn{6}{c}{Water} \\			
	\hline
	2.86 & 105480 & 0.134 & 1850 & 23.6 & 17.9 \\
	3.85 & 147660 & 0.173 & 2490 & 25.3 & 29.8 \\
	5.86 & 232810 & 0.249 & 3700 & 26.8 & 61.7 \\
	7.14 & 256290 & 0.304 & 4090 & 22.4 & 92.1 \\
	\hline
	\multicolumn{6}{c}{Carbopol 0.06\%} \\	
	\hline
	2.79 & 9430 & 0.161 & 200 & 25.4 & 25.9 \\
	3.80 & 19370 & 0.210 & 400 & 25.8 & 44.1 \\
	5.72 & 50670 & 0.296 & 970 & 26.6 & 87.8 \\
	\hline	
	\multicolumn{6}{c}{Carbopol 0.08\%} \\	
	\hline
	5.82 & 24740 & 0.306 & 490 & 26.9 & 93.4 \\
	\hline
	\multicolumn{6}{c}{Carbopol 0.10\%} \\
	\hline
	7.08 & 21000 & 0.373 & 420 & 26.7 & 139.0 \\
	\hline
\end{tabular} \caption{\fontsize{9}{9}\selectfont Parameters for water and Carbopol flow loop experiments.}
\label{parameters}
\end{table}


\section{Results and discussion: effect of Carbopol rheology} \label{Results1}

\subsection{Rheology and pressure measurements} \label{Results1_rheo}

\begin{figure}
	\centering
	\includegraphics[width=150mm]{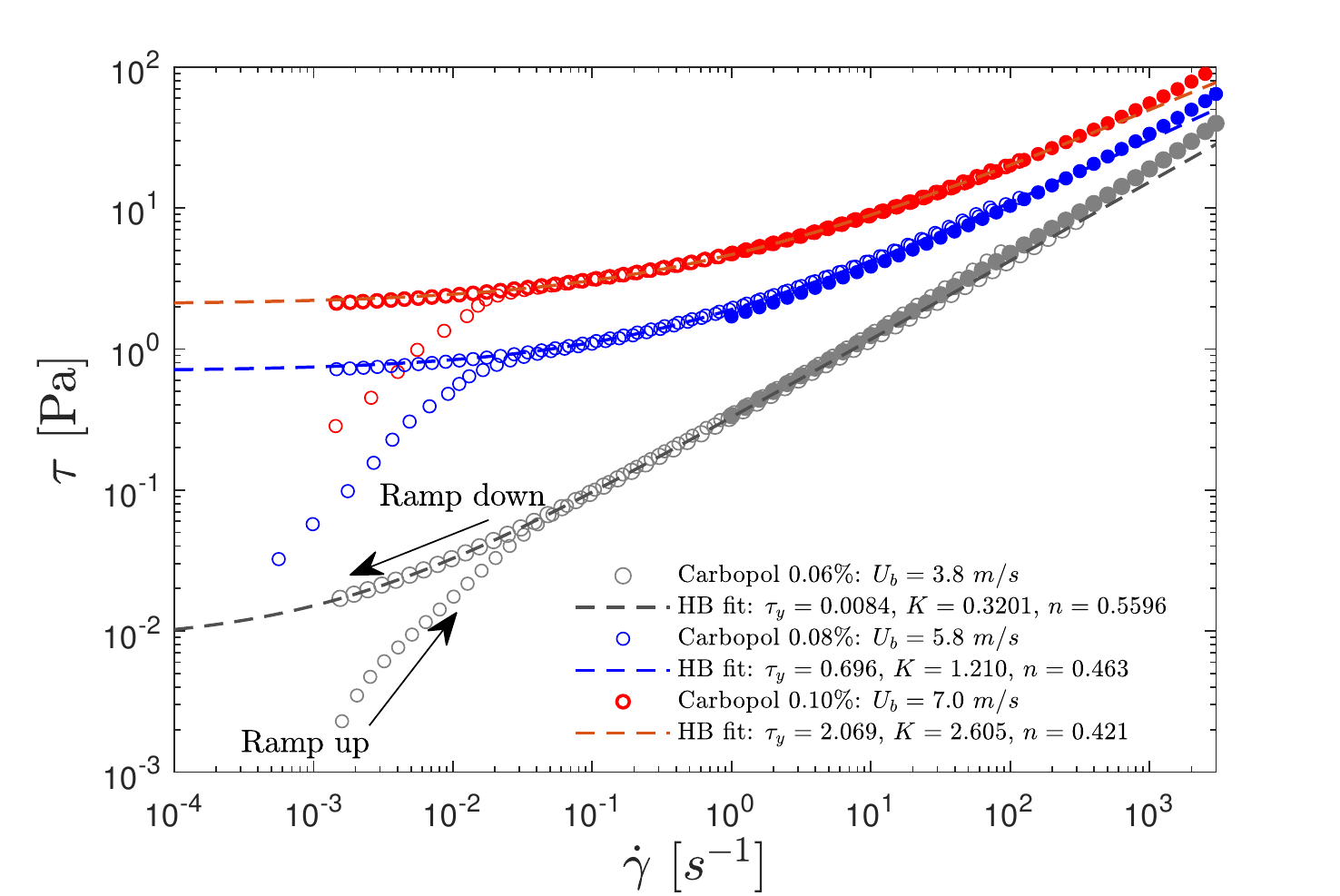}
	\caption{\fontsize{9}{9}\selectfont Flow curves of the Carbopol solutions used in flow loop. Empty symbols correspond to the ramp experiments, and filled symbols correspond to the steady state flow curves. The dashed lines represent fits to the Herschel-Bulkley model} 
	\label{flowcurves}
\end{figure}

To better contextualize the turbulence measurements, we first present the rheological characterization of the Carbopol solutions. \hyperref[flowcurves]{Figure~\ref*{flowcurves}} shows the ramp-up and ramp-down flow curves with samples from the three Carbopol solutions, collected from the tank after each LDA traverse. The ramp time was 3 minutes from a shear rate of $0.001$ to $200$ s$^{-1}$. All three Carbopol solutions exhibit yield stress and shear-thinning behaviour. The ramp down flow curves were then fitted to the Hershel-Bulkley model to estimate the dynamic yield stress, consistency index and power law index. There is a small deviation between the model fit and the steady shear experiments only at very high shear rates, over 1000 s$^{-1}$. No effects of inertia were observed in the rheological measurements due to the low gap between the parallel-plate geometry. The fitting parameters are also reported in \hyperref[flowcurves]{Figure~\ref*{flowcurves}}. As usual with Carbopol solutions, we observe hysteresis at low shear rates which is likely a consequence of the elasticity of the fluids near the yielding point \citep{balmforth2014}. The steady-shear-rate viscosity curves match the ramp-up/down curves quite well, with no discernible hysteresis at the high shear rates relevant for turbulent flows; thus, we consider that all Carbopol solutions shown in this work are non-thixotropic. 

\begin{figure}
	\centering
	\includegraphics[width=\textwidth]{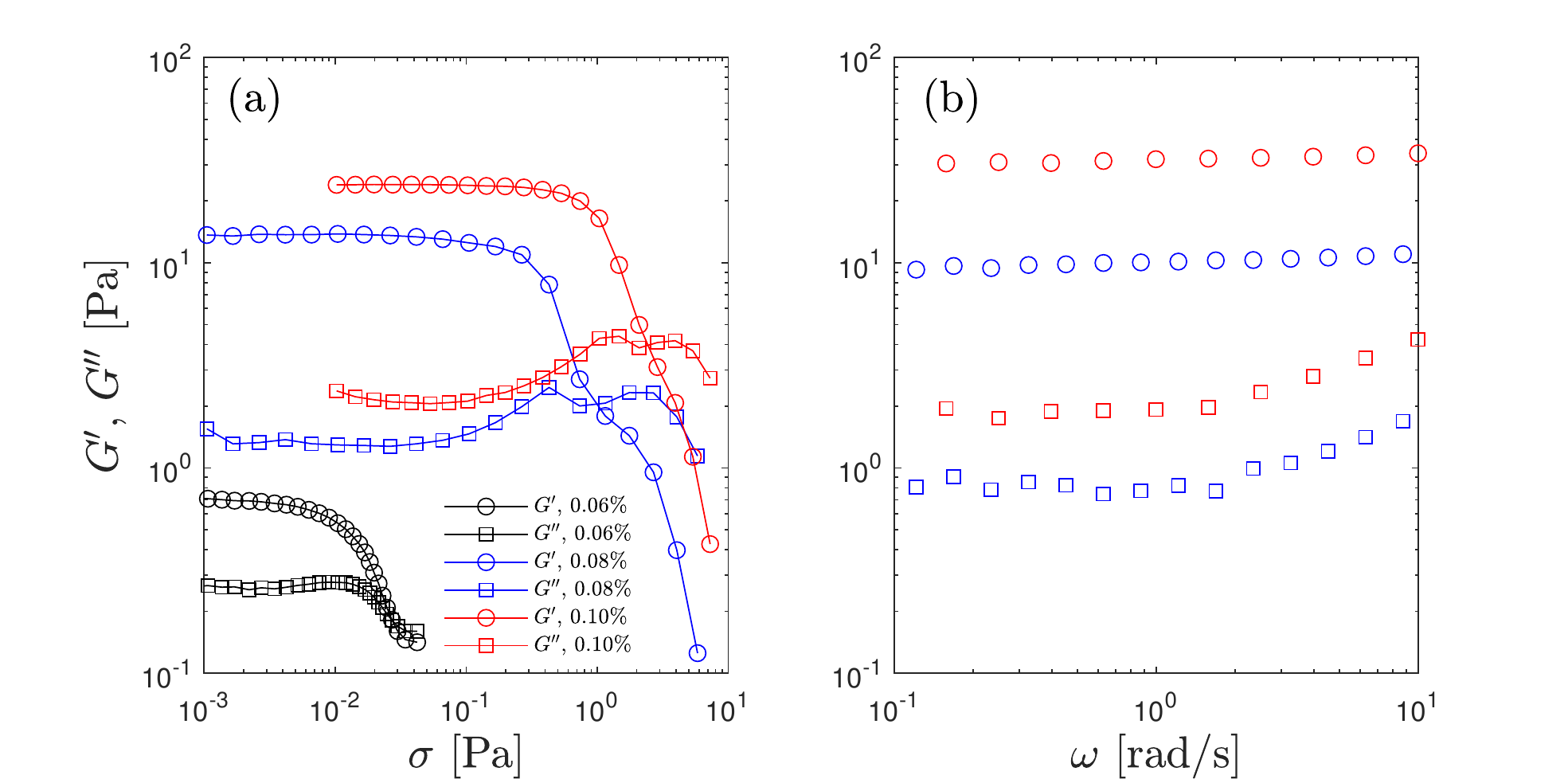}
	\caption{\fontsize{9}{9}\selectfont Stress amplitude sweep at $\omega = 3.14$ \textit{rad/s} (a) and small amplitude oscillatory shear measurements (b) at $\gamma = 0.1$ \% for Carbopol solutions.} 
	\label{carbopol_amp}
\end{figure}

We characterize the elastic properties of Carbopol by performing stress-controlled amplitude sweeps for each of the Carbopol solutions. The results are reported in \hyperref[carbopol_amp]{Figure~\ref*{carbopol_amp}} (a). The amplitude sweep results show that the elasticity is important in the linear viscoelastic region of constant $G'$ and $G''$ where the fluid is unyielded. The linear viscoelastic region falls below the yield stress which is of the same order of magnitude of the stress value at the crossover point between $G'$ and $G''$ \citep{dinkgreve2016}. Within the linear viscoelastic regime obtained from the amplitude sweeps, we perform strain-controlled, small amplitude oscillatory shear experiments (SAOS) with both 0.08\% and 0.10\% Carbopol solutions. SAOS measurements with the 0.06\% solution were not possible due to inertial effects. The results from \hyperref[carbopol_amp]{Figure~\ref*{carbopol_amp}} (b) show the usual gel-like behaviour of the solutions, which are characterized by a $G'$ plateau approximately independent of the frequency of oscillation $\omega$. $G''$ is also independent of the frequency below 1 \textit{rad/s}. We note here that we assume that the viscoelasticity in the non-linear regime with Carbopol solutions is small, according to other experimental works \citep{jossic2013, eslami2017}. We can confirm this statement by estimating the relaxation time of the Carbopol in the non-linear viscoelastic regime with first normal stress differences ($N_1$) measurements at steady shear with a cone-plate geometry. However, these experiments are often very challenging or impossible at the high shear rates in the boundary layer of turbulent flows, which can be over 1000 s$^{-1}$. Because of this reason, we did not perform measurements of $N_1$ in this work.

Friction factor measurements are useful to determine whether or not the experiments have reached the fully turbulent regime. From the pressure drop measurements, we compute the mean wall shear stress and the Fanning friction factor from the pipe flow of water and the three Carbopol solutions, according to the equation 
\begin{equation}
	f = \frac{2 \tau_w}{\rho U_b^2}.
\end{equation}

\begin{figure}
	\centering
	\includegraphics[width=80mm]{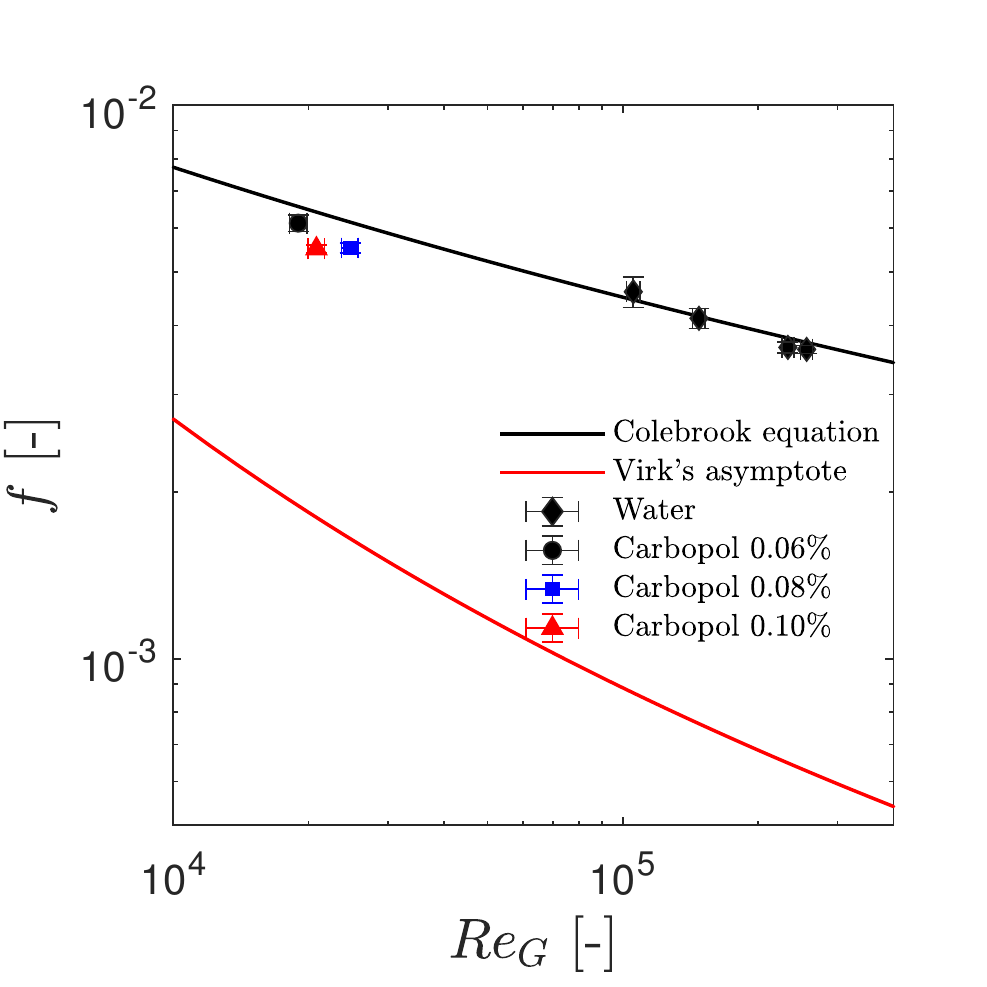}
	\caption{\fontsize{9}{9}\selectfont Measurements of the Fanning friction factor of the Carbopol solutions.} 
	\label{carbopol_ff}
\end{figure}

For reference, we compare our experimental friction factor results to the empirical Colebrook equation for the Fanning friction factor given by
 
\begin{equation}
	\frac{1}{ \sqrt{f} }  = -4.0 \, log \, \left( \frac{\epsilon/D_h}{3.7} + \frac{ 1.255 }{ Re_G \sqrt{f} } \right),
\end{equation}

\noindent which estimates the friction factor of turbulent Newtonian flows of $Re > 10^5$ with smooth inner surfaces (roughness $\epsilon$ = 0). We also show the asymptote of Virk \citep{virk1970},

\begin{equation}
	\frac{1}{ \sqrt{f} }  = 19.0 \, log \, \left( Re_G \, \sqrt{f} \right)  - 32.4,
\end{equation}

\noindent which predicts the state of maximum drag reduction in turbulent flows with viscoelastic polymer solutions. The friction factor results are plotted in \hyperref[carbopol_ff]{Figure~\ref*{carbopol_ff}} against the generalized Reynolds number.

The friction factor results for water nicely match the Colebrook equation. The Carbopol results show a limited drag reduction even at fully turbulent Reynolds numbers depicted here, when compared to the friction factor given by the Colebrook equation \textit{at the same generalized Reynolds number}. The small difference suggests that there is little viscoelasticity effect in the experiments shown in \hyperref[carbopol_ff]{Figure~\ref*{carbopol_ff}}. However, if we plot the friction factor of Carbopol as a function of the solvent (water) Reynolds numbers, we observe a drag increase with the addition of Carbopol at the same bulk velocities because of the increase in viscosity. This is due to the fact that a Carbopol solution is not a viscoelastic drag reducing fluid and can be considered to be an ideal viscoplastic fluid in turbulent flow conditions. DNS simulations with generalized Newtonian fluids such as those by \citet{singh2017a,singh2017b} are qualitatively similar.

\subsection{Turbulence statistics} \label{Results1_stats}

\begin{figure}
	\centering
	\includegraphics[width=100mm]{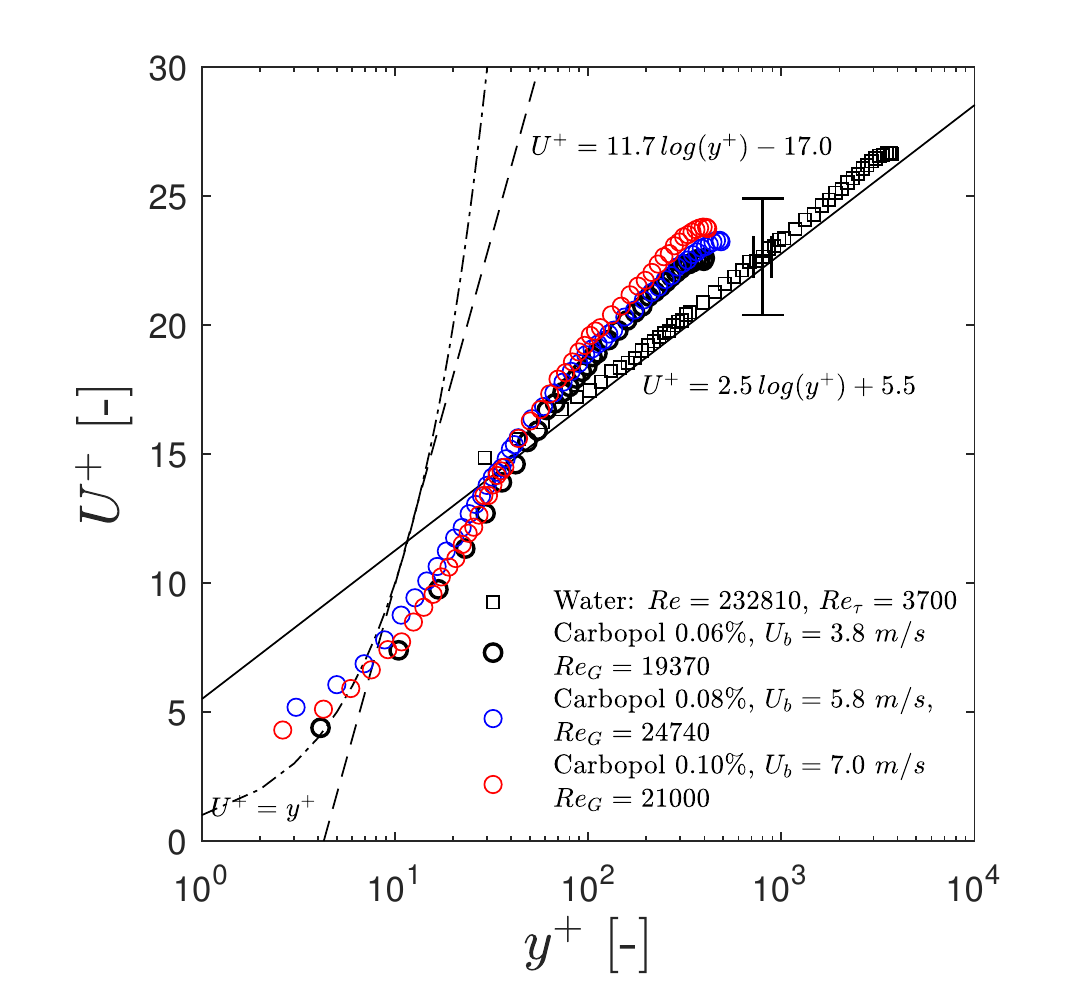}
	\caption{\fontsize{9}{9}\selectfont Velocity profiles of the turbulent flow of water and Carbopol solutions in wall units. The dot-dashed lines represent the viscous sublayer, the full lines represent the log law of the wall for turbulent flow of Newtonian fluids, and the dashed line shows Virk's asymptotic velocity profile for drag reducing polymer solutions. The error bar represents the variation of $u_{\tau}$ along the walls of the channel cross-section.}
	\label{carbopol_vel}
\end{figure}

Now we present the turbulence measurements of average velocities and Reynolds stresses of the flow of Carbopol from the LDA velocity measurements and compare those to Newtonian flows, both from our experiments with water and with DNS. For the analysis of LDA measurements, we employ the Reynolds decomposition of the Navier-Stokes equations \citep{pope2001}. The instantaneous streamwise velocity component $U$ is given by the sum of the mean streamwise velocity $\langle U \rangle$ and the fluctuations $u$: $U = \langle U \rangle + u$. The same decomposition is applied to the wall normal velocity component: $V = \langle V \rangle + v$. The profiles of the local time-averaged streamwise velocities normalized with the friction velocity ($U^+ = \langle U \rangle/ u_{\tau}$) are plotted against the wall normal position normalized by the wall unit ($y^+ = y/y^+_0$) in \hyperref[carbopol_vel]{Figure~\ref*{carbopol_vel}}. Note that the wall unit changes for each fluid formulation due to different effective viscosities at the wall. The velocity profiles, which were measured along the channel centre plane, are shown in \hyperref[carbopol_vel]{Figure~\ref*{carbopol_vel}}.

We observe good agreement between the experimental data set of water at $U_b = 5.8$ \textit{m/s} ($Re$ = 232810) and the log law profile. The velocity profiles of Carbopol show a slight upward shift in the log-layer when compared to water flow, consistent with experimental \citep{peixinho2005} results with Carbopol and numerical \citep{singh2017a,rosti2018} results with inelastic generalized Newtonian (GN) models for viscoplastic fluids. Therefore, the upward shift in our velocity profiles could be a consequence of shear thinning of the fluid, and not viscoelasticity, which appears to be significant only in the linear viscoelastic regime (or before the yielding transition). We also observe a very slight concentration dependence in the velocity profiles of each solution. The velocity profile of the 0.10\% solution is slightly steeper than the 0.06\% solution. This is likely due to the decrease of the power-law index $n$ with Carbopol concentration; in other words, an increase in shear-thinning effects. A similar result was been observed in the DNS results of \citep{rudman2004} for turbulent pipe flow of power-law fluids with decreasing $n$ values. The increase in the slope of the velocity profile agrees with the decrease in friction factor with concentration as shown in \hyperref[carbopol_ff]{Figure~\ref*{carbopol_ff}}. 

Regarding the aspect ratio of our channel, we did not perform experiments with the Carbopol in positions away form the channel centre plane because of fluid degradation concerns with the additional time required to perform these experiments. Nevertheless, we note that our flow cannot be approximated to a channel flow because of the 2:1 aspect ratio of the rectangular duct. To support our data, we show that the turbulence statistics of water are quite similar to what is calculated from Newtonian flow DNS at similar Reynolds number. Therefore, we believe this limitation in our experiment does not affect our main conclusions of this study. Another effect that is difficult to quantify is that only the mean wall shear stress is evaluated from the pressure measurements. Our channel has a 2:1 aspect ratio and we expect therefore to have significant variations in wall shear stress along the wall and into the corners, especially for a shear-thinning yield stress fluid. This results in $u_{\tau}$ varying $\sim 15\%$ across the width of channel. To clarify this effect of the geometry of our setup, the 15\% variation of $u_{\tau}$ across the channel is represented by an error bar in \hyperref[carbopol_vel]{Figure~\ref*{carbopol_vel}}. However, even with the differences in wall shear stress across the channel, the mean velocity profile of water shows good agreement with the Newtonian log-law profile, and the computation of $u_{\tau}$ with the mean wall shear stress is adequate. Note that if we do not consider the variation of mean wall shear stress in the channel, the estimated uncertainty in $U^+$ is near $3\%$, following the procedure of \citet{whalley2017}. 

\begin{figure}
	\centering
	\includegraphics[width=\textwidth]{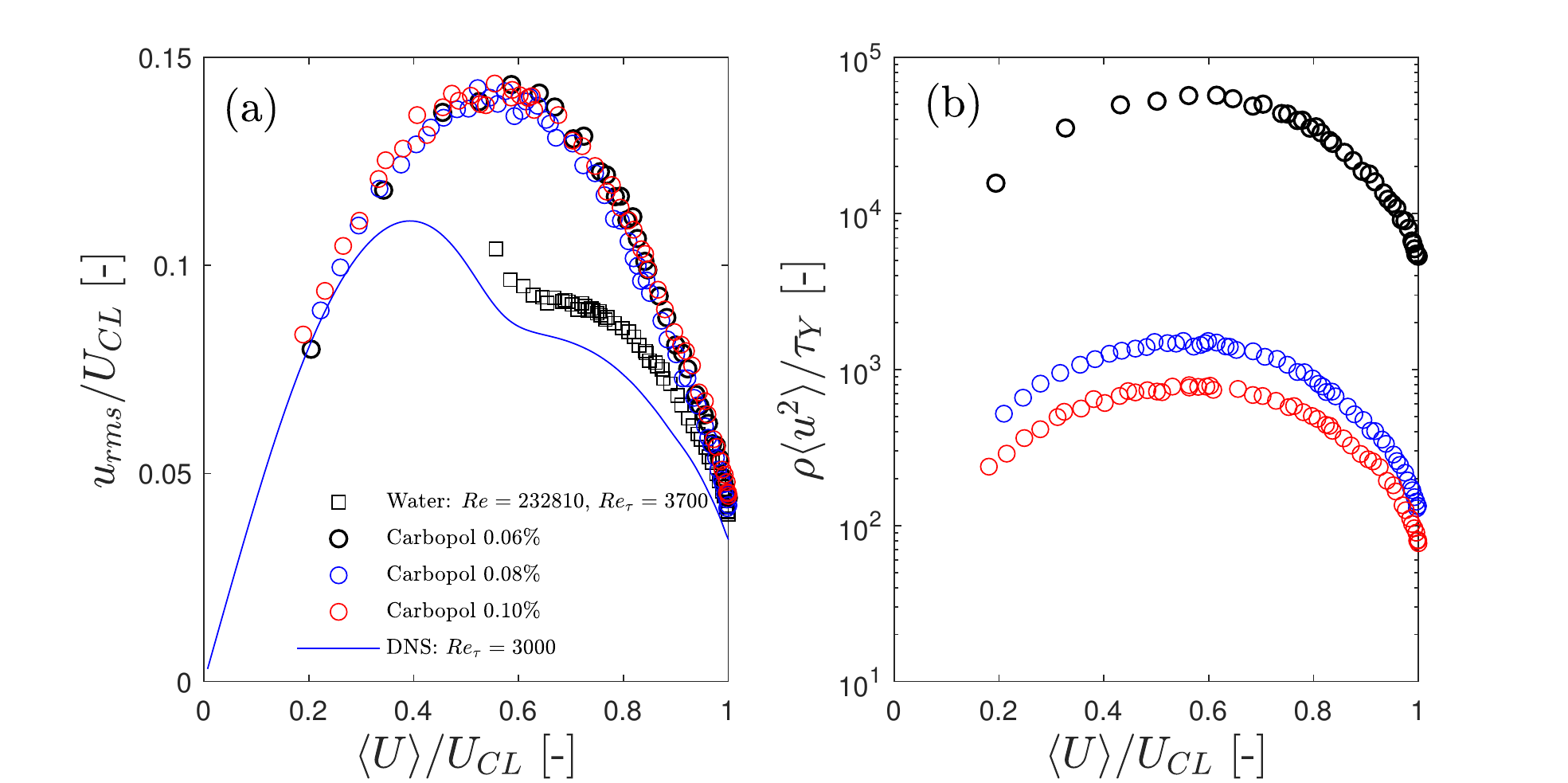}
	\caption{\fontsize{9}{9}\selectfont Streamwise turbulence intensities of the turbulent flow of water and Carbopol solutions normalized by the centreline velocity (a) and streamwise Reynolds stresses of the Carbopol solutions normalized by the yield stress (b), plotted against the local average velocity normalized by the centreline velocity.} 
	\label{carbopol_diagnostic}
\end{figure}

\hyperref[carbopol_diagnostic]{Figure~\ref*{carbopol_diagnostic}} (a) shows a diagnostic plot \citep{alfredsson2012}, where the root mean square of the velocity fluctuations normalized with the mean velocity measured at the centre of the channel $U_{CL}$, is plotted as a function of the mean local velocity normalized by the mean centreline velocity $\langle U \rangle/U_{CL}$. This is a useful representation of the turbulence intensities $u_{rms}$ (or $\sqrt{\langle u^2 \rangle}$) because it allows us to visualize the behaviour of the fluctuations without the wall position. We compare our experimental results with water to DNS results of Newtonian flows by \citet{ahn2015} at $Re_{\tau} = 3000$ (shown in dark blue lines) in a pipe. The reason we present DNS pipe flow results instead of channel flow is to indicate that even though our channel flow is three-dimensional due to the 2:1 aspect ratio, the experimental results with water in the centre plane of the channel (see \hyperref[LDA_schematic]{Figure~\ref*{LDA_schematic}}) agree, at least qualitatively. The Carbopol results are quite different from water. We notice a large increase in the turbulent intensities throughout most of the $y$-direction. The peak in turbulence production moves away from the wall, and is not dependent of the concentration of the Carbopol solution. We observe a good collapse of the Carbopol $u_{rms}$ data in the diagnostic plot, with no visible differences between concentrations. Although we cannot measure velocity in the buffer layer ($5 \leq y^+ \leq 30$) and viscous sub-layer ($y^+ < 5$) with water due to the very small boundary layer thickness at the high Reynolds numbers of our experiments, the data show good qualitative agreement with the Newtonian DNS at similar Reynolds number. \hyperref[carbopol_diagnostic]{Figure~\ref*{carbopol_diagnostic}} (b) indicates the streamwise Reynolds stresses are at least two orders of magnitude larger than the yield stress during all experiments presented in the velocity profiles of \hyperref[carbopol_vel]{Figure~\ref*{carbopol_vel}}. The calculation of the Bingham number $Bi = \tau_y/[K (U_b/h)^n]$ \citep{peixinho2005} also results in small values. According to \citep{guzel2009a}, this is shows that the Carbopol is fully yielded during all measurements performed here, and confirms that our experiments are indeed in the fully turbulent regime. Therefore, the results seem to point towards a negligible influence of the yield stress during fully turbulent flows of viscoplastic fluids.

\begin{figure}
	\centering
	\includegraphics[width=\textwidth]{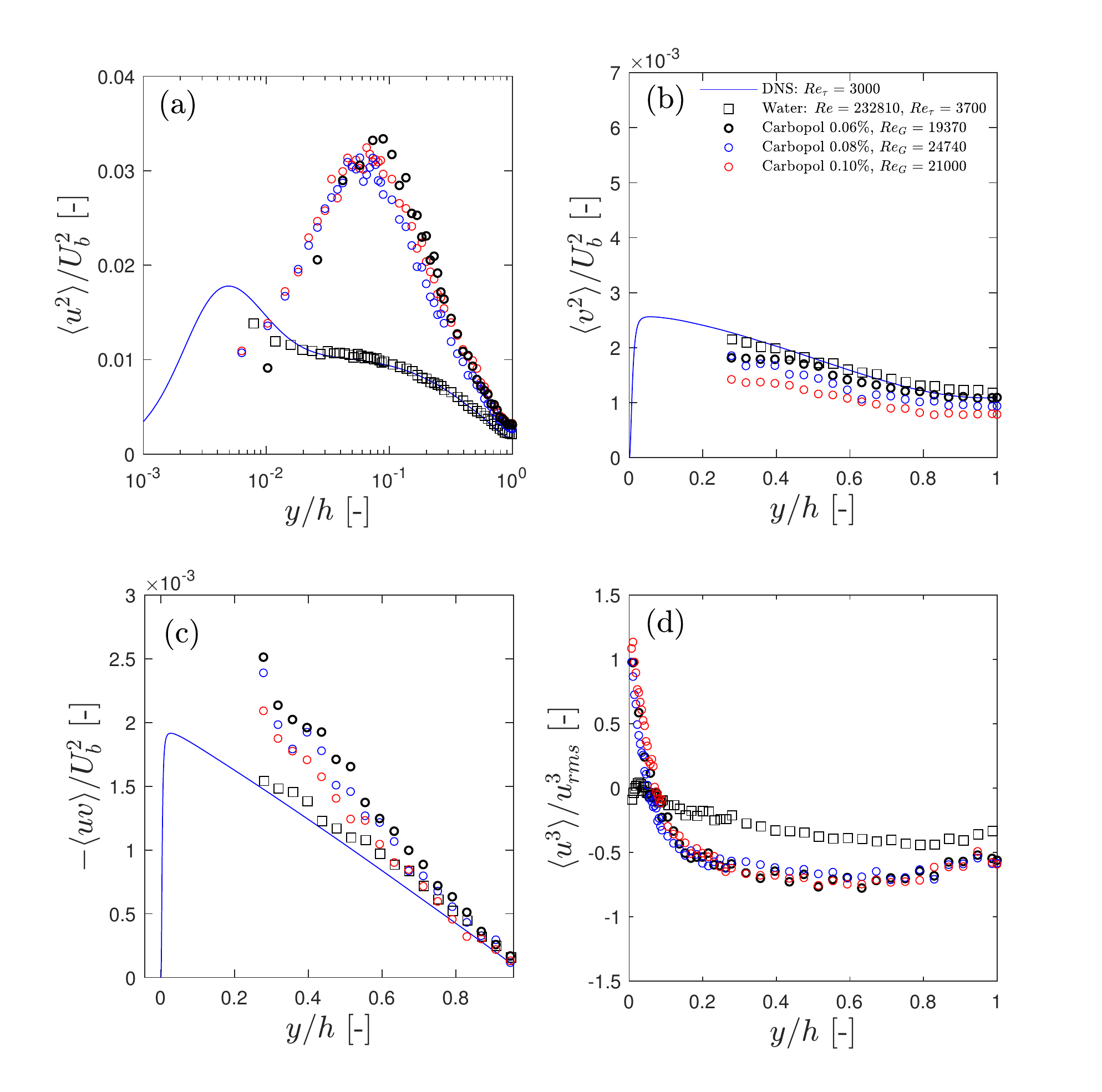}
	\caption{\fontsize{9}{9}\selectfont Streamwise Reynolds stresses $\langle u^2 \rangle$ (a), wall normal Reynolds stresses $\langle v^2 \rangle$ (b) and Reynolds shear stresses $-\langle uv \rangle$ (c) normalized by the bulk velocity $U_b^2$ and skewness $\langle u^3 \rangle/u_{rms}^3$ (d) of water and Carbopol solutions.} 
	\label{carbopol_stats}
\end{figure}

Reynolds stresses of water and Carbopol solutions are shown in \hyperref[carbopol_stats]{Figure~\ref*{carbopol_stats}} (a-c). All turbulent quantities in \hyperref[carbopol_stats]{Figure~\ref*{carbopol_stats}} (a-c) are normalized with $U_b$, which corresponds to the bulk velocity calculated from the flow meter during the experiments, except for $\langle u^3 \rangle$ which is normalized by $u_{rms}^3 = \langle u^2 \rangle^{3/2}$, resulting in the skewness in \hyperref[carbopol_stats]{Figure~\ref*{carbopol_stats}} (d). The streamwise turbulence intensities show a similar behaviour as the diagnostic plot except for the fact that we display it as a function of position this time. The peak in $\langle u^2 \rangle$ occurs near $y/h \sim 0.1$ for all Carbopol concentrations, and the fluctuations seem slightly larger for the 0.10\% Carbopol in comparison to the other concentrations. Overall the streamwise turbulence intensities are larger than for Newtonian fluids, and the decrease of the power-law index $n$ with Carbopol concentration does not seem to play a very large role here. These experimental observations agree qualitatively with the results from \citep{peixinho2005} and \citep{rosti2018}, although the increase in $\langle u^2 \rangle$ is rather small in their numerical simulations, when compared to experiment. The measurements of the wall normal turbulent intensities in \hyperref[carbopol_stats]{Figure~\ref*{carbopol_stats}} (b) are quite similar to Newtonian fluids, albeit at slightly smaller values overall. The effect of the parameter $n$ is more apparent in \hyperref[carbopol_stats]{Figure~\ref*{carbopol_stats}} (b) than (a), where the wall-normal fluctuations decrease with $n$. We note here that due to limitations of our LDA system, it is difficult to measure the $V$ component of velocity below $y/h \sim 0.28$. Nevertheless, our measurements still highlight the increase in anisotropy of the turbulent flow of Carbopol near the wall, indicated by the significant increase in $u$ fluctuations and slight decrease in $v$ fluctuations when compared to water. Turbulence anisotropy is also enhanced by larger concentrations of Carbopol, with decreasing $n$ values. 

Reynolds shear stresses -$\langle uv \rangle$ in Carbopol solutions, shown in \hyperref[carbopol_stats]{Figure~\ref*{carbopol_stats}} (c), seem to increase near the wall when compared to water, which is likely due to the large $\langle u^2 \rangle$ observed in \hyperref[carbopol_stats]{Figure~\ref*{carbopol_stats}} (a). Due to the impossibility of near wall measurements, we cannot be sure of the behaviour of the Reynolds shear stresses at lower values of $y/h$. However, from the $\langle u^2 \rangle$ results, we can speculate that the peak in Reynolds shear stresses might occur farther away from the wall than the Newtonian case. The values of -$\langle uv \rangle$ are also expected to be lower than water near the wall, similar to the trend seen in the $\langle u^2 \rangle$ results. The observation that all values converge near the core is also interesting. The Reynolds and viscous shear stresses vanish in the turbulent core, and we speculate that the yield stress effect might play a role at suppressing turbulent fluctuations. Additionally, we note that in case of water flow, all results displayed here agree with the available DNS results for pipe flow for $Re_\tau = 3000$ when scaled with $U_b$. 

The skewness values shown in \hyperref[carbopol_stats]{Figure~\ref*{carbopol_stats}} (d) depict changes in the near wall turbulence dynamics in the flows of Carbopol when compared to Newtonian fluids. An increase in positive skewness near the wall indicates more intensity in sweeping motions, or high speed fluid moving towards the wall ($u > 0, v < 0$) \citep{wallace1972} when compared to water, and further away from the wall we observe an inflection point near $y/h = 0.07$, after which the skewness of the fluctuations become negative. The negative skewness indicates enhanced ejection motions, characterized by low speed fluid moving away from the wall ($u < 0, v > 0$) \citep{wallace1972}, which supports the results from \hyperref[carbopol_stats]{Figure~\ref*{carbopol_stats}} (a) where the increase in streamwise fluctuations denote more production of turbulence \citep{mohammadtabar2017}. Overall, \hyperref[carbopol_stats]{Figure~\ref*{carbopol_stats}} (a-d) highlights changes in Reynolds stresses due to the addition of Carbopol, but there is not a large difference between the three concentrations, due to the rather narrow power law index $n$ range in the formulated solutions.

\begin{figure}
	\centering
	\includegraphics[width=\textwidth]{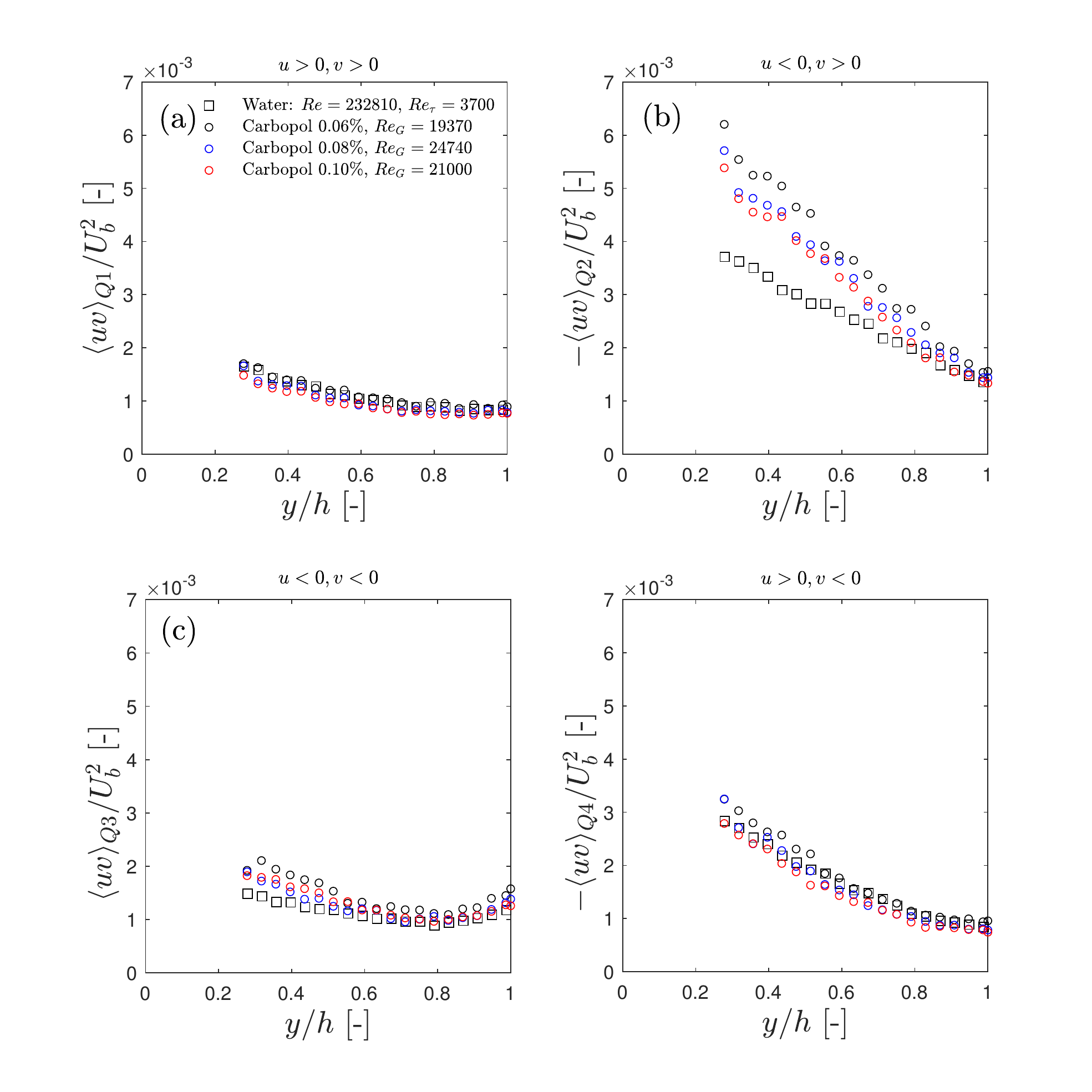}
	\caption{\fontsize{9}{9}\selectfont Conditional averages of Reynolds stresses of the turbulent flow of water and Carbopol solutions plotted in the four quadrants: $\langle uv \rangle_{Q1}$ (a), $\langle uv \rangle_{Q2}$ (b), $\langle uv \rangle_{Q3}$ (c) and $\langle uv \rangle_{Q4}$ (d).} 
	\label{carbopol_uv_cond}
\end{figure}

\hyperref[carbopol_uv_cond]{Figure~\ref*{carbopol_uv_cond}} presents conditional averages of the Reynolds shear stresses with respect of $\langle uv \rangle$ with respect to the sign of the velocity fluctuations $u$ and $v$: $u > 0, v > 0$: Q1 events or outward interactions); $u < 0, v > 0$: Q2 events or ejections; $u < 0, v < 0$: Q3 events or wallward interactions and $u > 0, v < 0$: Q4 events or sweeps \citep{wallace1972}. The largest contributions towards the Reynolds shear stresses come from Q2 events or ejection motions, and Q4 events or sweep motions \citep{adrian2007} as seen from \hyperref[carbopol_uv_cond]{Figure~\ref*{carbopol_uv_cond}} (b) and (d). In the particular case of Carbopol solutions, both Q1, Q3 and Q4 seem relatively unaltered when compared to water, whereas Q2 events in \hyperref[carbopol_uv_cond]{Figure~\ref*{carbopol_uv_cond}} (b) and (c), are the most affected. Specifically for \hyperref[carbopol_uv_cond]{Figure~\ref*{carbopol_uv_cond}} (b), the rheological changes in the fluid impart a large increase in ejection motions for all fluids studied here. This is connected to the large enhancement in streamwise turbulence intensities seen previously. 

\begin{figure}
	\centering
	\includegraphics[width=\textwidth]{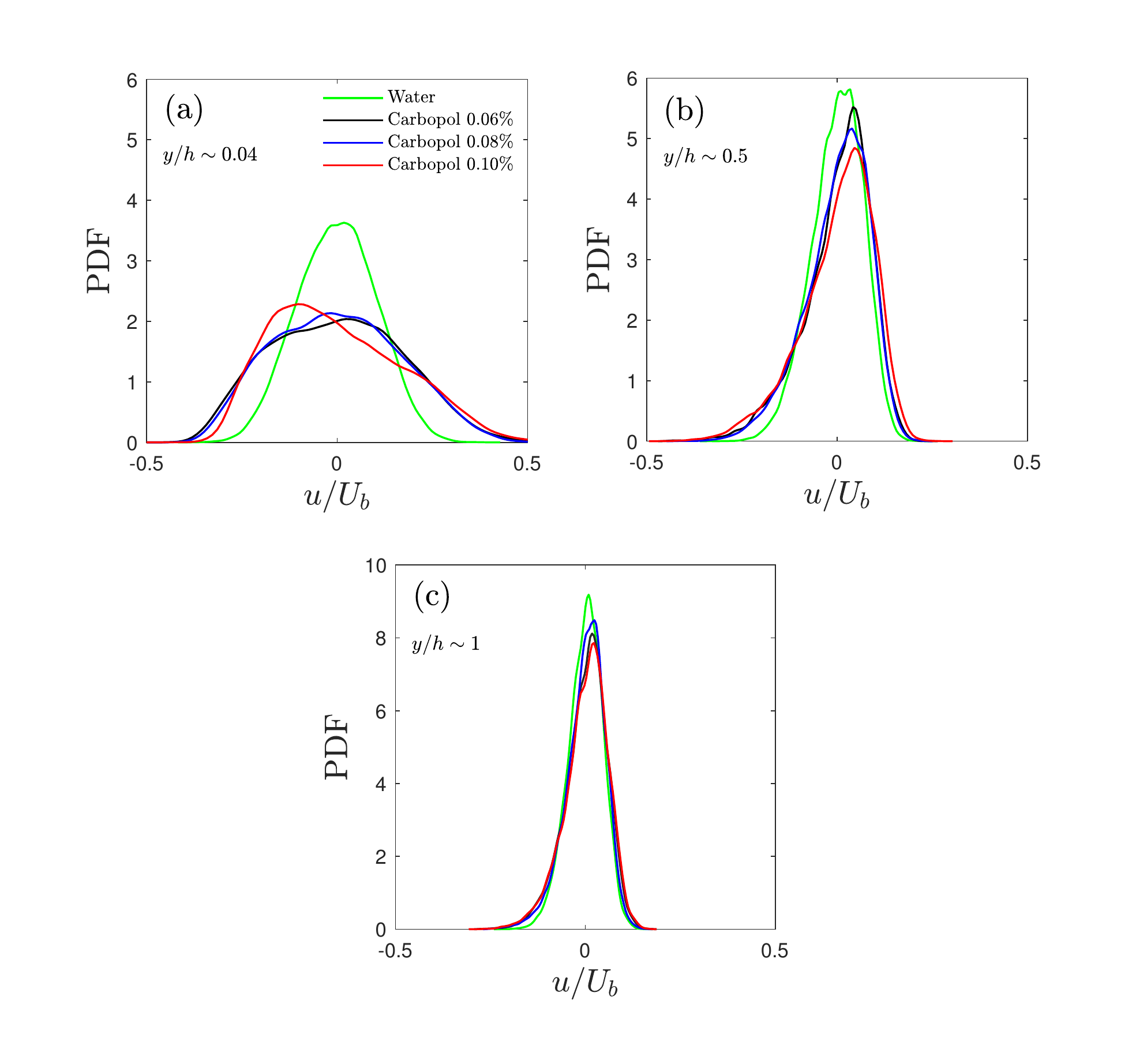}
	\caption{\fontsize{9}{9}\selectfont Probability density functions of $u/U_b$ of water and Carbopol solutions. Image (a) corresponds to $y/h \sim 0.04$ while image (b) shows $y/h \sim 0.5$} 
	\label{carbopol_pdf}
\end{figure}

We also show the probability density functions of the streamwise velocity fluctuations in \hyperref[carbopol_pdf]{Figure~\ref*{carbopol_pdf}} to highlight the difference in dynamics very close to the wall and near the core. Near the buffer layer, at $y/h \sim 0.04$ in \hyperref[carbopol_pdf]{Figure~\ref*{carbopol_pdf}} (a), we observe larger probabilities of large $u$ velocity fluctuations with Carbopol than water. The narrower pdf of water also indicates a larger probability of smaller $u$ velocity fluctuations. \hyperref[carbopol_pdf]{Figure~\ref*{carbopol_pdf}} (b) shows the probability density function (pdf) of $u$ velocity fluctuations at $y/h \sim 0.5$, where the difference between Carbopol and water is small, although this difference increases somewhat with the concentration. In this case, the pdf of the Carbopol solutions is slightly more skewed towards negative values, as previously shown by \hyperref[carbopol_stats]{Figure~\ref*{carbopol_stats}} (d). Also, the probability of small $u$ velocity fluctuations reduces slightly with the concentration of Carbopol. 

Before moving on to the next section, we summarize the main findings from the turbulence statistics of different Carbopol solutions. The main observation is that, for high $Re_G$ turbulence, the yield stress is insignificant relative to the Reynolds stresses. Therefore, similarly to previous studies such as \citep{dodge1959,maleki2016,rudman2004, singh2017b}, only the parameter $n$, or the amount the fluid shear thins, seems to be of most influence here. Indeed, from the perspective of dimensional analysis, where $\tau_y/\tau_w \ll 1$, only $n$ should affect the flow. The effects of $n$ are observed as an increase in the velocity profile when scaled with inner units, and a decrease in wall-normal and Reynolds shear stresses , $\langle v^2 \rangle$ and -$\langle uv \rangle$. Conversely, the streamwise Reynolds stresses $\langle u^2 \rangle$ increase quite significantly, and we conclude that the main consequence of Carbopol addition in the Reynolds stresses to enhance streamwise turbulent structures, making the flow more anisotropic. This is likely due to the more pronounced viscosity differences across the channel with larger shear-thinning effects - the larger viscosities in the core dampen the transfer of momentum to the wall, leading to a decrease in $\langle v^2 \rangle$ and -$\langle uv \rangle$ \citep{gavrilov2016, singh2017b}, and thus increased $\langle u^2 \rangle$. However, our range of $n$ is quite narrow to observe large changes in statistics. Since we cover $0.42 \leq n \leq 0.56$, our results across all three Carbopol formulations are quite similar. Numerical simulations allow coverage of a much wider range of parameters. We note that the experimental conditions would not allow much smaller values of $n$ to be achieved, since we need to increase the Carbopol concentration to decrease $n$. The resultant increase in viscosity and $\tau_y$ would make reaching turbulence more difficult.

\subsection{Power spectral densities} \label{Results1_spectra}


We investigate the one-dimensional power spectral densities (PSDs), represented by $E_{uu}$ for streamwise velocity fluctuations $u(t)$ and $E_{vv}$ for wall normal velocity fluctuations $v(t)$. We compare energy spectra results with water and Carbopol solutions in different channel positions. The power spectral density is an estimate of the energy distribution throughout the frequency range, i.e.~$E_{uu} \propto |u_f(f)|^2$, where $u_f(f)$ is the Fourier transform of $u(t)$, for a given $y/h$ position. The frequency data $f$ is then converted into wavenumber space by $k_x = 2 \pi f / \langle U \rangle$, and made dimensionless multiplied by with the boundary layer thickness $h$, or channel half-height, for fully developed channel flow. This is possible by using Taylor's hypothesis, which has been widely used for LDA experiments \citep{pope2001, warholic1999influence}. We consider that the hypothesis is acceptable if the root mean square of the velocity fluctuations $u$ and $v$ is less than 20\% of the mean local velocity $\langle U \rangle$. The PSD is made dimensionless dividing by $\nu_s U_b$, where $\nu_s$ is the kinematic viscosity of the solvent. The normalization is to aid comparison between different fluids.

We employ linear interpolation of the velocity-time signal to obtain equally spaced data points. The interpolation frequency is the average data rate of the experiments (similar to \citep{toonder1997}) which varies between 3000 and 4000 \textit{Hz} for the $U$ component, and between 2000 and 4000 \textit{Hz} for the $V$ component. The limitation of this method is that the interpolation acts as a filter to the data signal at high frequencies, and because of this, we cut off our spectral results at a maximum frequency of approximately 1/4 of the total data rate, which is a reasonable approximation as shown by \citet{ramond2000}, for instance. Even without the high frequency results, we are able to estimate the inertial range spectra in our water results.  

\begin{figure}
	\centering
	\includegraphics[width=\textwidth]{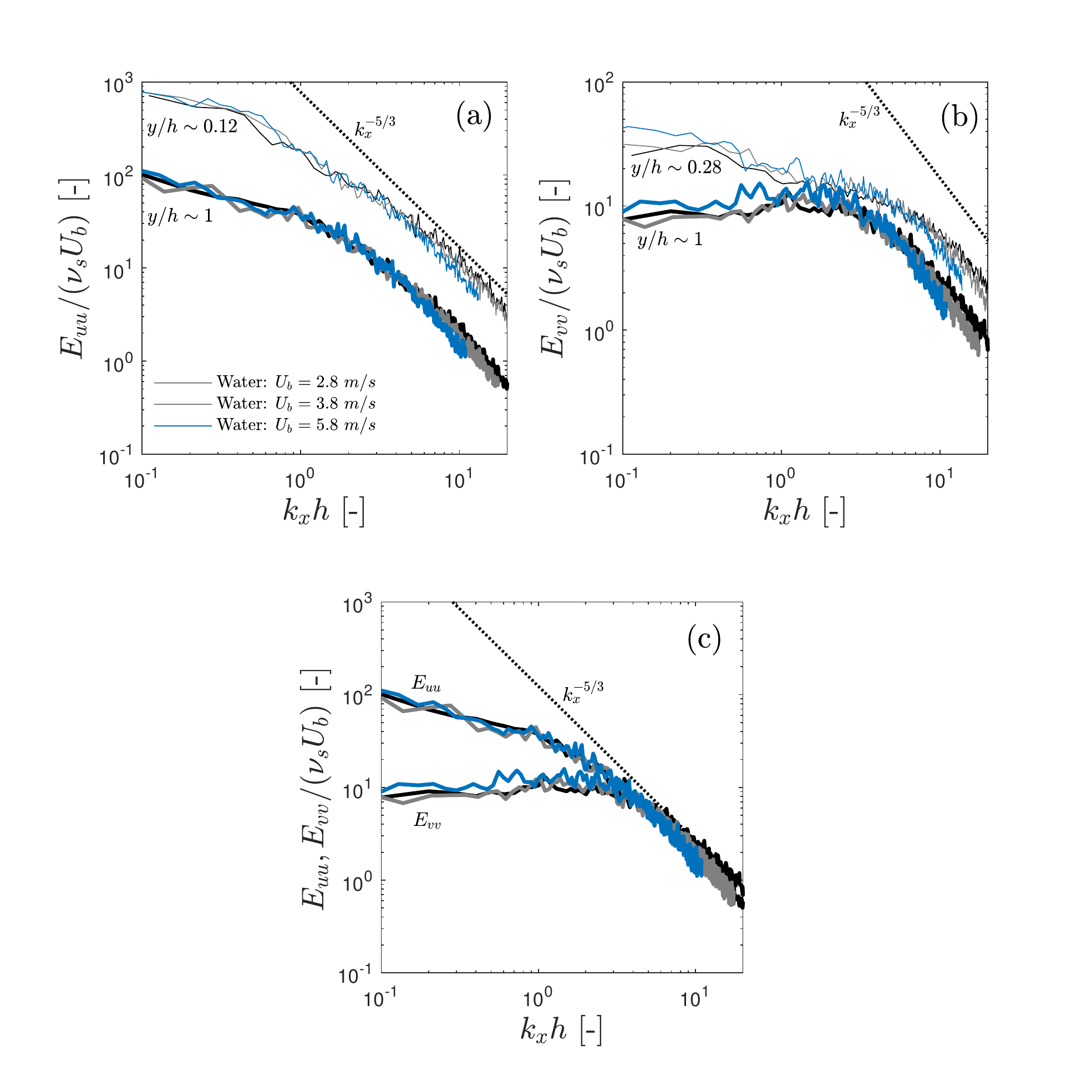}
	\caption{\fontsize{9}{9}\selectfont Power spectral densities of velocity fluctuations during the flow of water. (a) shows the PSDs of the streamwise velocity fluctuations at $y/h \sim 0.12$ (thin lines) and at the centre plane at $y/h \sim 1$ (thick lines). The $k_x^{-5/3}$ scaling is shown by dotted lines.  (b) shows the PSDs of the wall normal velocity fluctuations at $y/h \sim 0.28$ (thin lines) and at the centreline at $y/h \sim 1$ (thick lines). Image (c) shows the PSDs of streamwise ($E_{uu}$, thin lines) and wall normal velocity fluctuations ($E_{vv}$, thick lines) at the centreline $y/h \sim 1$.} 
	\label{water_spec1}
\end{figure}

As an initial benchmark, we show the PSDs of the $u$ fluctuations, given by $E_{uu}$, in \hyperref[water_spec1]{Figure~\ref*{water_spec1}} (a) and $v$ fluctuations  given by $E_{vv}$ in \hyperref[water_spec1]{Figure~\ref*{water_spec1}} (b) for water at three distinct Reynolds numbers at various $y/h$ positions specified in the figures. The dotted lines represent the wavenumber scaling for the inertial range of the turbulent energy cascade $k_x^{-5/3}$, where the energy transfer from large to small eddies depends only on the energy dissipation rate \citep{pope2001}. By means of scaling with the bulk velocity $U_b$ and the kinematic viscosity $\nu_s$, there is a very good collapse, both for $E_{uu}$ and $E_{vv}$. We observe that, for both positions shown in \hyperref[water_spec1]{Figure~\ref*{water_spec1}} (a), energy decay from small wavenumbers (representative of large eddies) to large wavenumbers (representative of small eddies) follow the $k_x^{-5/3}$ scaling quite closely for $k_x h > 5$. With \hyperref[water_spec1]{Figure~\ref*{water_spec1}}, we can establish that the experimental results are able to provide the energy spectra in the inertial range at reasonable wavenumbers. A similar conclusion is drawn from \hyperref[water_spec1]{Figure~\ref*{water_spec1}} (b), where the energy content from $v$ fluctuations also falls with the same $k_x^{-5/3}$ typically expected from Newtonian fluids at high $Re$. \hyperref[water_spec1]{Figure~\ref*{water_spec1}} (c) presents $E_{uu}$ and $E_{vv}$ at the centre of the channel. In this figure we observe a typical behaviour of isotropic turbulence (at least considering the measurement plane for our LDA results) in the centre of the channel, in which both $E_{uu}$ and $E_{vv}$ collapse in the high wavenumber range, also with approximately a $k_x^{-5/3}$ scaling \citep{cerbus2020}. 

\begin{figure}
	\centering
	\includegraphics[width=\textwidth]{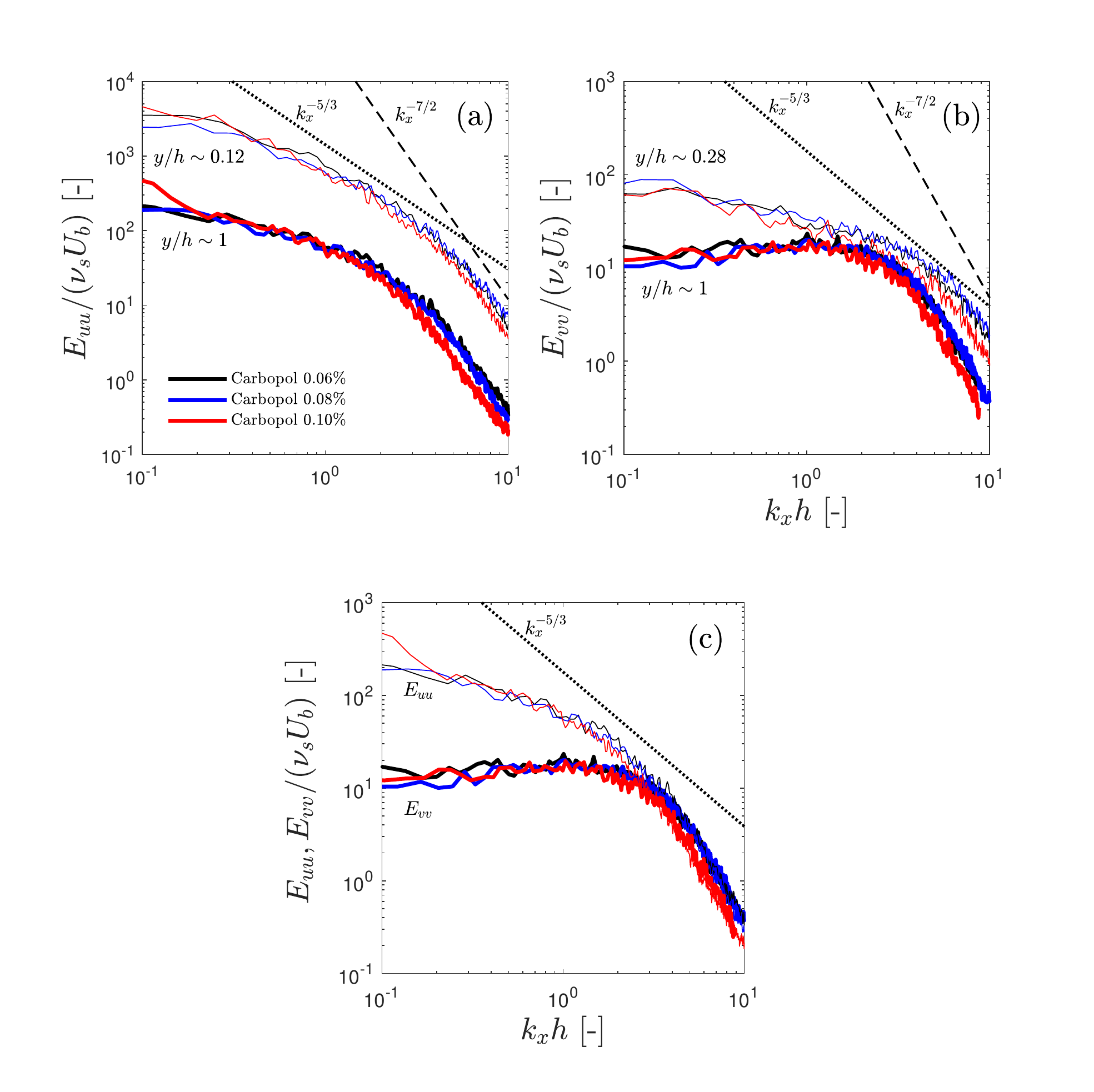}
	\caption{\fontsize{9}{9}\selectfont Power spectral densities of velocity fluctuations during the flow of Carbopol solutions in concentrations of 0.06 (black), 0.08 (blue) and 0.10\% (red). Image (a) shows the PSDs of the streamwise velocity fluctuations at $y/h \sim 0.12$ (thin lines) and at the centre plane at $y/h \sim 1$ (thick lines). The $k_x^{-5/3}$ scaling is shown by the dotted line, and the $k_x^{-7/2}$ scaling is shown by the dashed line. Image (b) shows the PSDs of the wall normal velocity fluctuations at $y/h \sim 0.28$ (thin, lines) and at the centre plane at $y/h \sim 1$ (thick lines). The plot (c) shows the PSDs of streamwise ($E_{uu}$, thin lines) and wall normal velocity fluctuations ($E_{vv}$, thick lines) at the centre plane ($y/h \sim 1$).} 
	\label{carbopol_spec1}
\end{figure}

The PSDs of streamwise velocity fluctuations with the addition of Carbopol are shown in \hyperref[carbopol_spec1]{Figure~\ref*{carbopol_spec1}}, for similar $Re_G$ conditions. In \hyperref[carbopol_spec1]{Figure~\ref*{carbopol_spec1}} (a), the curves for different Carbopol concentrations are quite close to each other by scaling with $U_b$ and $\nu_s$ at each $y/h$ position. Interestingly, the scaling of energy for higher wavenumbers ($k_x h > 5$) for the Carbopol measurements is close to $k_x^{-7/2}$ instead of the usual $k_x^{-5/3}$ for the inertial range in Newtonian fluids, indicating a larger power decay for higher wavenumbers (small scales). The energy drop off at large wavenumbers also appears to be independent of Carbopol concentration and position in the channel, at least for the range of parameters investigated here. The energy spectra decay of $k_x^{-5/3}$ can be observed in the PSDs of Carbopol, but in a lower range of $1 \leq k_x h \leq 4$ approximately, which is a consequence of low Reynolds numbers. The theoretical studies by \citet{anbarlooei2017} and \citet{anbarlooei2018} have proposed that Kolmogorov's $k_x^{-5/3}$ scaling for the energy spectra in the inertial range is valid for viscoplastic fluids, to enable scaling expressions for the wall shear stress and thus to compute the friction factor in pipe flows. An interesting observation from this section is that the PSDs appear to confirm their statement because the $k_x^{-5/3}$ power law is still observed, albeit for different $k_x h$ values than water. 

The PSDs of wall-normal velocity fluctuations shown in \hyperref[carbopol_spec1]{Figure~\ref*{carbopol_spec1}} (b) also follow the $k_x^{-7/2}$ scaling at large wavenumbers approximately, but unlike \hyperref[carbopol_spec1]{Figure~\ref*{carbopol_spec1}} (a) the $k_x^{-5/3}$ scaling is not as apparent at intermediate wavenumbers. The good collapse of the spectra results in \hyperref[carbopol_spec1]{Figure~\ref*{carbopol_spec1}} (a) is also seen in \hyperref[carbopol_spec1]{Figure~\ref*{carbopol_spec1}} (b). The energy content of $v$ fluctuations of 0.1\% Carbopol is slightly lower overall, which correlates to the lower values of $\langle v^2 \rangle$ observed previously. \hyperref[carbopol_spec1]{Figure~\ref*{carbopol_spec1}} (c) shows that the Carbopol solution is also isotropic in the centreline in the high wavenumber range, albeit with the spectra scaling to the power law of $k_x^{-7/2}$. It is interesting to observe isotropy at the centreline at high $k_x$ range, even though there is a clear increase in turbulent anisotropy near $y/h \sim 0.1$ due to the large peak in $\langle u^2 \rangle$. Summarizing the results from \hyperref[carbopol_spec1]{Figure~\ref*{carbopol_spec1}}, the PSDs for all three Carbopol formulations investigated here are quite similar to each other. The similarity in the PSDs are expected since the values of streamwise Reynolds stresses $\langle u^2 \rangle$ were quite similar across all concentrations of Carbopol solutions (or values of the power-law index $n$) investigated. The values of wall-normal Reynolds stresses $\langle v^2 \rangle$ also did not change significantly. This means that the PSDs are not very sensitive to changes in the rheology of the fluid, not unlike the results observed in the turbulence statistics. We can expect that, similarly to what was observed in the wall-normal Reynolds stresses results, $\langle v^2 \rangle$ decreases with $n$, so it is likely that the energy content decreases as well. DNS results show that $\langle u^2 \rangle$ increases with $n$ \citep{rudman2004, singh2017b}, so following the same trend the energy spectra should also increase. However, without a wider range for the power-law index $n$, it remains difficult to experimentally investigate its effect. 

\begin{figure}
	\centering
	\includegraphics[width=\textwidth]{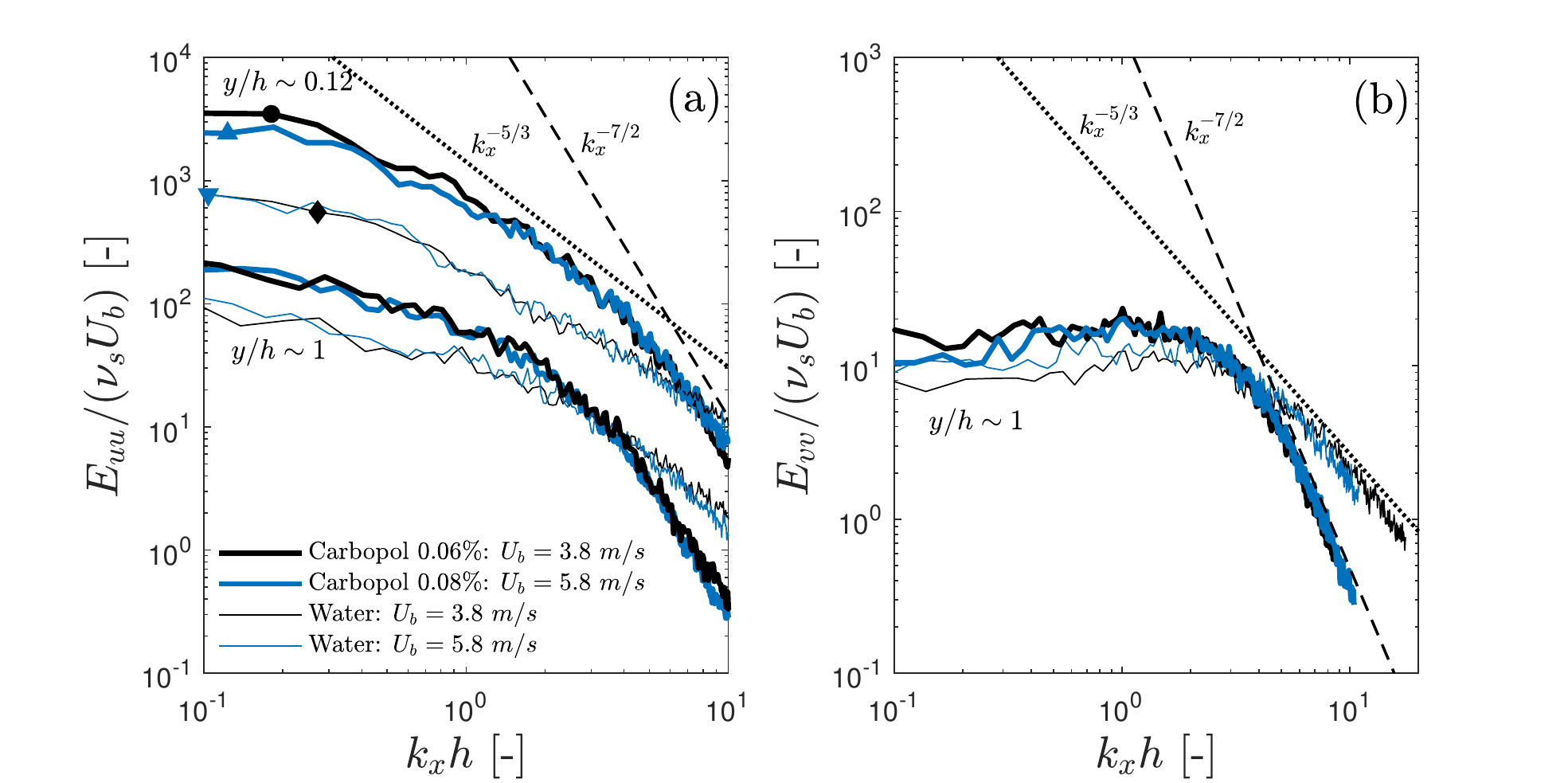}
	\caption{\fontsize{9}{9}\selectfont Power spectral densities of streamwise (a) and wall normal (b) velocity fluctuations during the flow of water and Carbopol solutions. Streamwise velocity fluctuations in (a) were measured at $y/h \sim 0.12$ (lines with symbols) which is the closest position to the wall where the Taylor's hypothesis is valid and $y/h \sim 1$. Wall normal velocity fluctuations in (b) were measured at $y/h \sim 1$.} 
	\label{water_carbopol_spec1}
\end{figure}

We compare the PSDs of both water and Carbopol at the same $U_b$ values in \hyperref[water_carbopol_spec1]{Figure~\ref*{water_carbopol_spec1}} in two positions in the channel centre plane. Closer to the wall in \hyperref[water_carbopol_spec1]{Figure~\ref*{water_carbopol_spec1}} (a), the addition of Carbopol to the flow enhances the energy content in small wavenumbers, or larger eddy length scale, when compared to water. This happens in both measurement positions shown. The consequence of this effect is the increase of streamwise fluctuations in the velocimetry measurements of \hyperref[carbopol_stats]{Figure~\ref*{carbopol_stats}} (a). As noted previously, the energy content in $k_x h > 5$ decreases as $k_x^{-7/2}$ for the Carbopol solutions, and $k_x^{-5/3}$ for water. The PSDs of the $v$ component fluctuations of Carbopol in \hyperref[water_carbopol_spec1]{Figure~\ref*{water_carbopol_spec1}} (b) are compared to the water results. Similarly to the $E_{uu}$ results, the energy at low wavenumbers is enhanced with the addition of Carbopol, while the energy content at high wavenumbers is lower. The $k_x^{-7/2}$ power law scale is also observed in the Carbopol results, as a consequence of less energy contained in $k_xh > 4$.

\begin{figure}
	\centering
	\includegraphics[width=\textwidth]{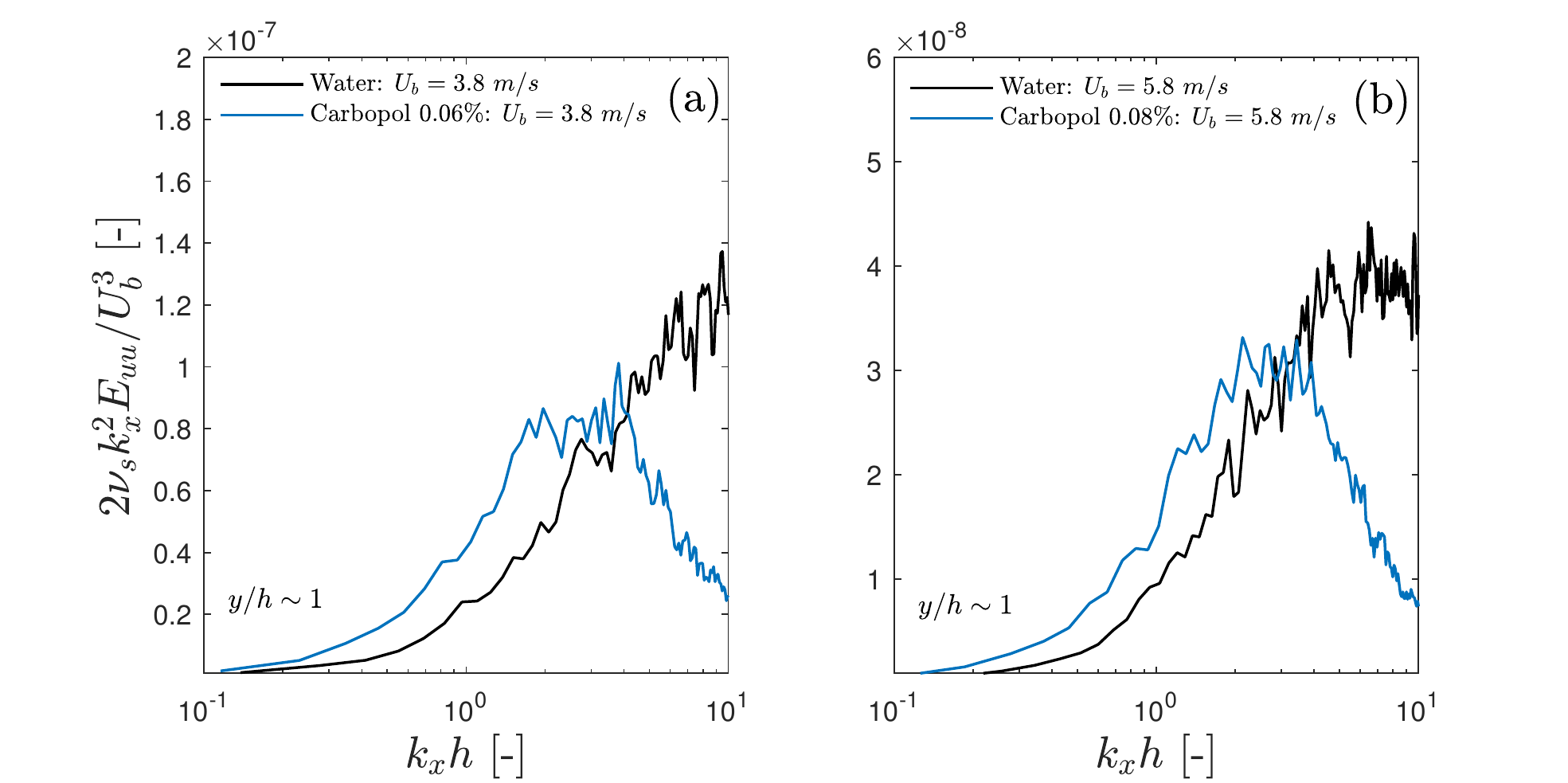}
	\caption{\fontsize{9}{9}\selectfont Estimation of the dissipation spectra for water and the 0.06\% Carbopol solution at $U_b = 3.8$ \textit{m/s} (a), and water and the 0.08\% Carbopol solution at $U_b = 5.8$ \textit{m/s} (b) from the respective streamwise PSDs, measured at $y/h \sim 1$.} 
	\label{water_carbopol_dissip}
\end{figure}

To further investigate the steeper slope in the high wavenumber range in the Carbopol solutions, we plot the dissipation spectra (which can be estimated by $2 \nu \int_{0}^{\infty} k^2 E_{uu}(k) dk$), normalized by $U_b^3$, in \hyperref[water_carbopol_dissip]{Figure~\ref*{water_carbopol_dissip}}. As with the PSD results, we cut off the dissipation spectra plot at $k_xh = 10$ to avoid aliased results due to the uneven data rate of the LDA. We observe that, for the same bulk velocities, the majority of the energy dissipation in Carbopol occurs at larger scales than with water. There is a peak at $k_xh = 2$ and a decrease in dissipation at $k_xh > 4$ in both \hyperref[water_carbopol_dissip]{Figure~\ref*{water_carbopol_dissip}} (a) and (b). The decay in dissipation is not well-resolved in the water results due to data rate limitations, but we see that a large amount of energy dissipation happens at smaller scales than in the Carbopol solutions.

\section{Results and discussion: effect of the Reynolds number} \label{Results2}

\begin{figure}
	\centering
	\includegraphics[width=\textwidth]{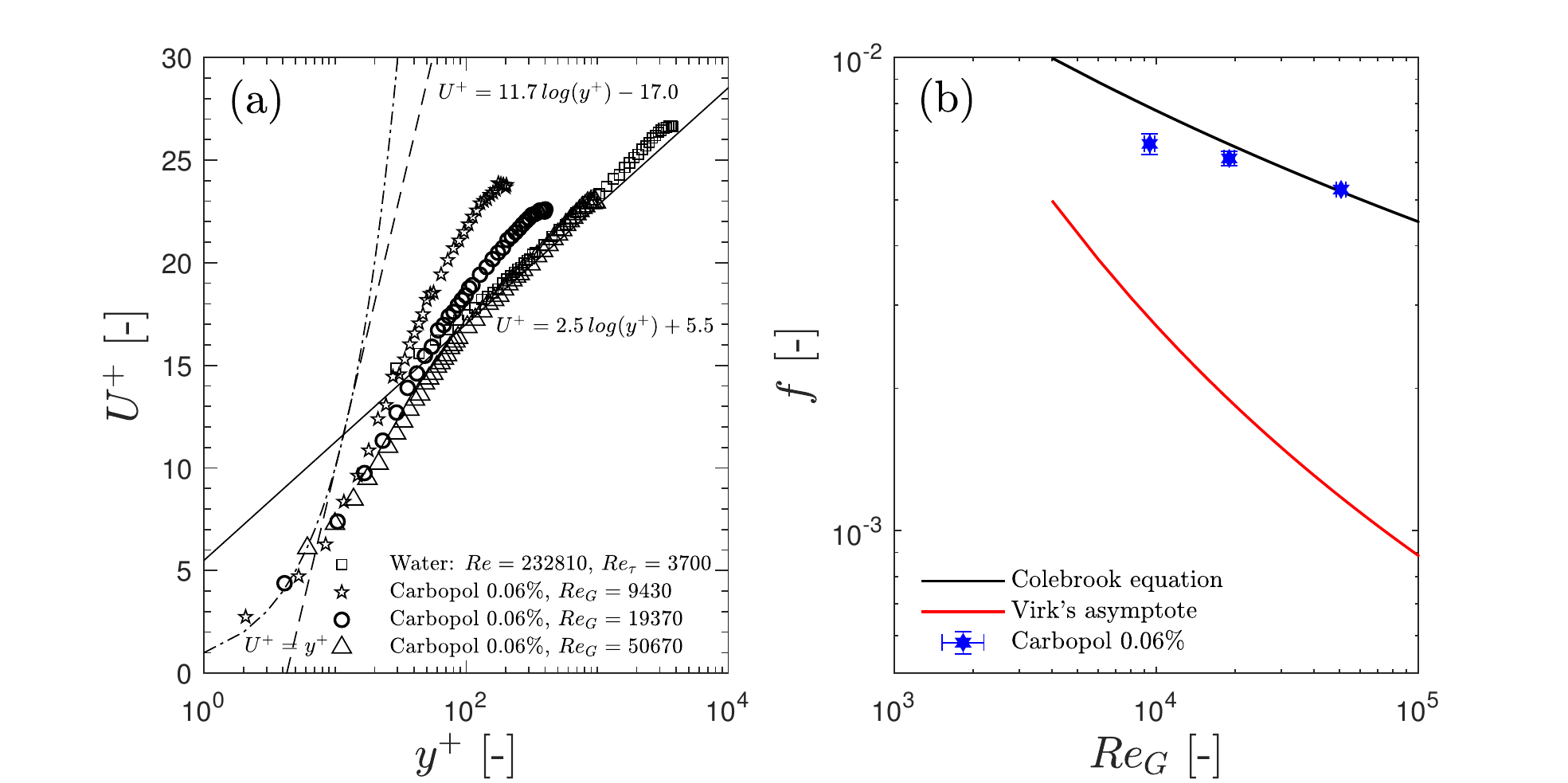}
	\caption{\fontsize{9}{9}\selectfont Mean velocity profiles normalized in wall units (a) and friction factor (b) during turbulent flow of 0.06\% Carbopol solutions. In plot (a), the dot-dashed lines represent the viscous sublayer, the full lines represent the log law of the wall for turbulent flow of Newtonian fluids, and the dashed line shows Virk's asymptotic velocity profile for drag reducing polymer solutions.} 
	\label{carbopol_vel2}
\end{figure}

In this section we investigate the flow of Carbopol at 0.06\% concentration at different $Re_G$ values, which range from weakly turbulent at $Re_G = 9430$ to strongly turbulent $Re_G =50670$. We remind the reader that the full experimental parameters are listed in \hyperref[parameters]{Table~\ref*{parameters}}. We consider the rheology of the fluid at all three Reynolds numbers investigated here to be the same as presented in \hyperref[flowcurves]{Figure~\ref*{flowcurves}}. Indeed, rheological tests with samples taken after experiments at different Reynolds numbers have shown little effect of degradation (not shown). The velocity profiles normalized by the friction velocity $U^+ = \langle U \rangle/u_{\tau}$ plotted against the wall-normal positions normalized by the wall unit $y^+ = y/y_0^+$ are presented in \hyperref[carbopol_vel2]{Figure~\ref*{carbopol_vel2}} (a) for each $Re_G$ values, and the friction factor results, calculated similarly to \hyperref[carbopol_ff]{Figure~\ref*{carbopol_ff}}, are shown in \hyperref[carbopol_vel2]{Figure~\ref*{carbopol_vel2}} (b).

The velocity profiles in \hyperref[carbopol_vel2]{Figure~\ref*{carbopol_vel2}} (a) show an interesting dependence in $Re_G$. At a low $Re_G$ value of 9430, the velocity profile is somewhat similar to the turbulent flow of viscoelastic, drag reducing (DR) fluids \citep{white2008, escudier2009}, where there is an upturn in the velocity profile near the buffer layer. This is also seen in the friction factor plot as a function of $Re_g$ in \hyperref[carbopol_vel2]{Figure~\ref*{carbopol_vel2}} (b), where the Carbopol friction factor is lower than the Newtonian case at $Re_G = 9430$, and moves towards the Colebrook friction factor line as $Re_G$ increases. \citet{escudier1996} observed similar DR-like results in pipe flow of a Laponite solution, which is a thixotropic viscoplastic fluid. However, their flow was clearly transitional at $Re_G = 3400$, whereas our Reynolds number is quite a bit higher. As we increase the Reynolds number, the flow seems to approach a Newtonian-like velocity profile, specifically at $Re_G = 50670$ where the velocity profile of Carbopol 0.06\% matches the log-law profile for Newtonian fluids. Therefore, the turbulent flow of Carbopol appears to become increasingly Newtonian-like as $Re_G$ is increased.

\begin{figure}
	\centering
	\includegraphics[width=\textwidth]{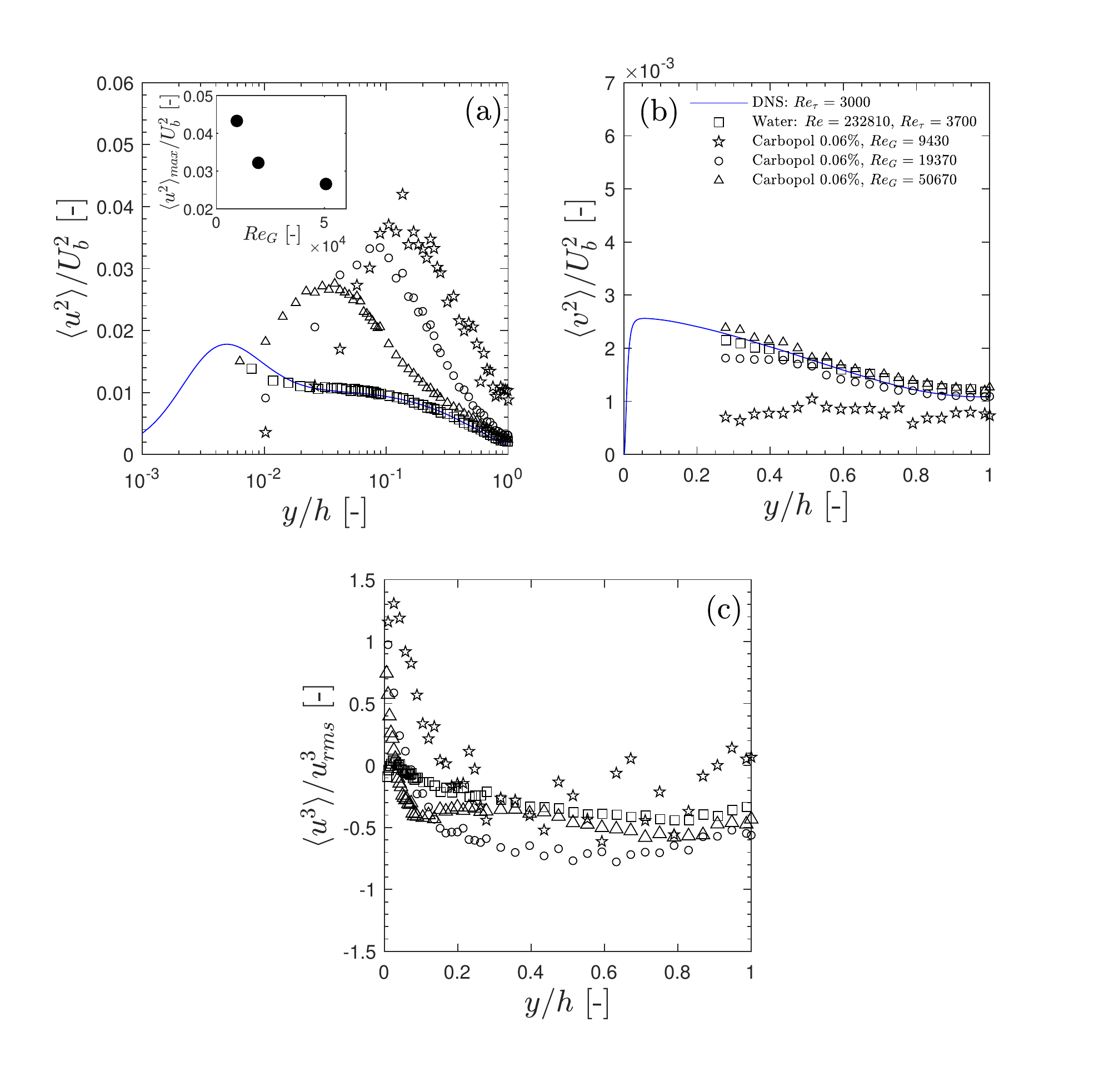}
	\caption{\fontsize{9}{9}\selectfont Streamwise Reynolds stresses $\langle u^2 \rangle$ (a), wall normal Reynolds stresses $\langle v^2 \rangle$ (b) normalized by the bulk velocity $U_b^2$, and skewness $\langle u^3 \rangle/u_{rms}^3$ (c) of water and Carbopol solutions.} 
	\label{carbopol_stats2}
\end{figure}

The flow statistics are presented in \hyperref[carbopol_stats2]{Figure~\ref*{carbopol_stats2}}. We note a significant change in streamwise Reynolds stresses with $Re_G$ in \hyperref[carbopol_stats2]{Figure~\ref*{carbopol_stats2}}(a). As we increase the Reynolds number, the peak in $\langle u^2 \rangle$ moves closer to the wall, and approaches the Newtonian values near the turbulent core as $Re_G$ is increased. It seems that for a very large Reynolds number, it is reasonable to expect the $\langle u^2 \rangle$ profile to further resemble the Newtonian in a larger range of $y/h$ values. Considering the $\langle v^2 \rangle$ results in \hyperref[carbopol_stats2]{Figure~\ref*{carbopol_stats2}}(b), a similar assessment can be made as the Carbopol values approach the Newtonian profiles as $Re_G$ increases. The values of $\langle v^2 \rangle$ are also quite low at $Re_G = 9430$, which correlates to larger $U^+$ in the velocity profiles in \hyperref[carbopol_vel2]{Figure~\ref*{carbopol_vel2}} (a). This effect of $Re_G$ in $\langle u^2 \rangle$ and $\langle v^2 \rangle$ is qualitatively similar to simulations by \citet{singh2017b}. As $Re_G$ increases, the effects of the polymer additive are observed ever closer to the wall. It is important to note that the Carbopol used in the present work does not appear to be viscoelastic according to data from \hyperref[Results1]{Section~\ref*{Results1}}, and the evidence provided by the experiments and numerical simulations of GN fluids \citep{rudman2004, singh2017b} seem to point out to shear-thinning effects only. The Reynolds number affects the skewness of the $u$ fluctuations as seen in \hyperref[carbopol_vel2]{Figure~\ref*{carbopol_vel2}} (c). As $Re_G$ increases, regions of positive skewness are observed closer to the wall. In addition, the skewness profile for the $Re_G = 50670$ case is quite similar to the water measurements, from $y/h \sim 0.4$.

It is difficult to arrive to a conclusive explanation for the $Re_G$ dependence of turbulence statistics. A similar result has been observed in simulations of turbulent channel flows of Bingham fluids in \citet{rosti2018}. They state that the fluid becomes Newtonian-like as the Bingham number, or the ratio of yield stress to viscous stress, decreases. By increasing $Re_G$ we effectively decrease the Bingham number of our flows as well. However, the range of $Re_G$ values in our experiments are already quite high so this Reynolds number dependence is more likely an effect of diminishing shear thinning effects due to the very large shear rates near the wall. As $Re_G$ increases, so do the shear rates, which may lead to the approach to an asymptotic viscosity plateau $\eta_{\infty}$ value for very high shear rates, similar to a Carreau-Yasuda fluid. At such high shear rates, the fluid's viscosity might not change as much compared to lower shear rates. However, rheometry limitations prevent us to estimate the viscosity at very high shear rates due to inertia and low-gap errors with the parallel-plate geometry.

\begin{figure}
	\centering
	\includegraphics[width=\textwidth]{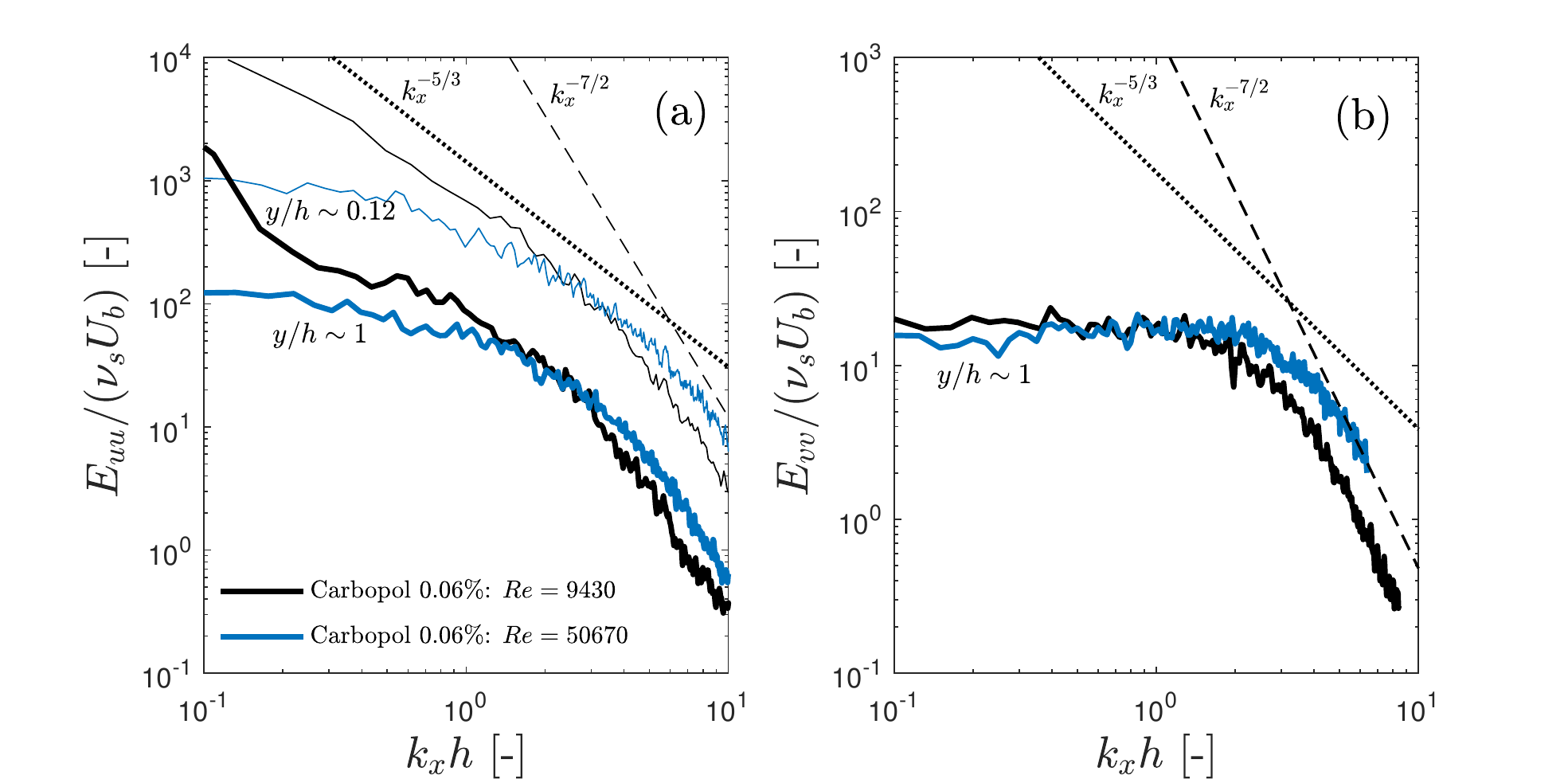}
	\caption{\fontsize{9}{9}\selectfont Power spectral densities of streamwise velocity fluctuations (a) and a comparison of streamwise and wall normal velocity fluctuations during the flow of Carbopol at 0.06\% concentration during two experimental conditions: $Re_G = 9430$ and $Re_G = 50670$. Streamwise velocity fluctuations in (a) were measured at $y/h \sim 0.12$ (thin lines) which is the closest position to the wall where the Taylor's hypothesis is valid, and the centreline of the channel $y/h \sim 1$ (thick lines). Measurements in (b) were performed at $y/h \sim 1$.} 
	\label{carbopol600_spec1}
\end{figure}

We investigate the effects of the Reynolds number on the PSDs of 0.06\% Carbopol solutions in \hyperref[carbopol600_spec1]{Figure~\ref*{carbopol600_spec1}}. Considering the PSDs of streamwise velocity fluctuations $u$ in \hyperref[carbopol600_spec1]{Figure~\ref*{carbopol600_spec1}} (a), we notice that the spectra seem to better resemble $k_x^{-5/3}$ decay as $Re_G$ increases, and is especially noticeable at $Re_G = 50670$ at $y/h \sim 0.12$, but changes are somewhat subtle in the high wavenumber range. The $E_{vv}$ energy content appears to increase with $Re_G$ at the centreline for $k_x > 1$. Then, larger $Re_G$ suggests an approach towards a Newtonian-like energy spectra in $E_{uu}$, with lower energy at small wavenumbers and an extended wavenumber range for $k_x^{-5/3}$, more evident for $y/h \sim 0.12$. These results suggest a diminished importance of the non-Newtonian rheology in the turbulence dynamics with large $Re_G$, although the effects of Carbopol such as increased anisotropy and large $\langle u^2 \rangle$ persist even at a very high $Re_G$ near the wall. Regarding the $k_x^{-7/2}$ drop in power, considering that the energy cascade from large to small eddies is supposed to be independent of viscosity in the inertial range \citep{pope2001}, we can hypothesise that the lower inertial effects in the turbulent flow of Carbopol might narrow the $k_x^{-5/3}$ wavenumber range. Then, the $k_x^{-7/2}$ may be a consequence of the energy decay due to dissipation at larger scales in the Carbopol solutions than water. Another reason could be increased elastic effects in the small (high frequency) length scales in the energy cascade, specifically. This explanation was proposed in \citet{presti2000} to justify the Reynolds number dependence of turbulent flows of Carbopol. In summary, the effect of Reynolds number in the turbulent flow of Carbopol appears to be the decrease of shear-thinning effects, both in the mean velocity profile and in the Reynolds stresses. This is more evident approach of the wall-normal Reynolds stresses $\langle v^2 \rangle$ to the Newtonian values. The $k_x^{-5/3}$ range in $E_{uu}$ is extended to smaller scales as $Re_G$ increases, again further resembling the water measurements. We can assume that with increasing shear rates due to the increase in $Re_G$, the effect of shear-thinning is diminished as an viscosity plateau $\eta_{\infty}$ is approached. 

\section{Conclusion} \label{Conclusion}

We performed an experimental investigation of the turbulent flows of Carbopol solutions with LDA measurements in a recirculating flow loop setup, with the main focus on both the rheology effect with similar $Re_G$, and the effect of different $Re_G$ with a constant fluid formulation. Three concentrations of Carbopol were investigated: 0.06\%, 0.08\% and 0.10\%. For comparison and benchmarking puruposes, we also presented turbulence data for the flow of water. All measurements are performed in the fully turbulent regime within a 2:1 channel, and the Carbopol is assumed to be fully yielded in all experiments.

We investigated the effect of shear-thinning of Carbopol solutions in the flow field and Reynolds stresses. The addition of Carbopol to water provides an slightly increased slope of the velocity profile $U^+(y^+)$ in comparison to the Newtonian (log law) case. The streamwise Reynolds stresses $\langle u^2 \rangle$ in Carbopol increase by a large amount when compared to water, but the effect of rheological changes between each Carbopol formulation was negligible in $\langle u^2 \rangle$. The wall-normal Reynolds stress component $\langle v^2 \rangle$ was affected more noticeably by changes in rheology, with the values of $\langle v^2 \rangle$ decreasing with concentration. This result suggests large anisotropy in turbulence structures near the wall due to an enhancement in streamwise turbulent structures. Overall, to observe a more significant effect of rheological parameters in Reynolds stresses, a wider range of formulations have to be studied. 

The increase in $\langle u^2 \rangle$ with the Carbopol addition in comparison to water is reflected in the $E_{uu}$ by an enhanced energy content at low wavenumber, when compared to the flow of water at the same $U_b$. The power spectral densities reveal that the energy scales with a $k_x^{-7/2}$ power law for large wavenumbers during Carbopol flows. This is in contrast to the $k_x^{-5/3}$ power law encountered in water in the same wavenumber range of $k_x h > 5$. The $k_x^{-5/3}$ scaling does not disappear with the addition of Carbopol, but happens in a lower range than water, near $1 \leq k_x h \leq 4$. The $k_x^{-7/2}$ slope is also seen in the $E_{vv}$ spectra. We can speculate that the $k_x^{-7/2}$ scaling in the high wavenumber range is due to decreased inertial effects in the turbulent flow of Carbopol solutions, since the Reynolds numbers are much lower than in water flows, with a narrower inertial range wavenumbers and with dissipation at larger length scales than water. Another possible explanation could be elasticity effects of the Carbopol, which could be significant only at small scales and high frequencies. Richardson's energy cascade theory predicts that the energy transfer in the inertial range is independent of viscosity \citep{pope2001}, and this statement supports our interpretation. Nevertheless, the turbulent spectra could be investigated in more detail, and we plan to carry out experiments in a circular cross-section pipe in a future study. 

The effect of the Reynolds number during the flow of Carbopol 0.06\% reveal interesting turbulence dynamics. At $Re_G = 9430$ the velocity profile and turbulent statistics reveal a large increase in the slope of $U^+(y^+)$, and smaller $\langle  v^2 \rangle$ quantities. Conversely, at the high Reynolds number case $Re_G = 50670$, the flow becomes somewhat similar to a Newtonian flow in aspects such as the velocity profile $U^+(y^+)$ matching the Newtonian log law, and streamwise Reynolds stresses also approaching Newtonian values near the core. We believe that this is due to the decrease of shear thinning effects in the turbulence statistics near the wall, where the shear rates are very high. However, the turbulence spectra showed little change in $k_x$ power-law scaling at the three $Re_G$ studied, but the energy content was slightly altered as $Re_G$ increased. We expect that the results presented in this work can aid in comparisons to turbulent flow experiments with other fluids and DNS simulations, especially when investigating the effects of shear-thinning rheology to the flow field.

\section*{Acknowledgments} \label{Ackn}
This research was made possible by research funding from Schlumberger and NSERC under the CRD program, project 505549-16. Experimental infrastructure was funded by the Canada Foundation for Innovation and the BC Knowledge Fund, grant number CFI JELF 36069. This funding is gratefully acknowledged. We also acknowledge support from the University of British Columbia 4 Year Fellowship PhD scholarship program (R.S.M.).

\bibliography{Results_carbopol}

\end{document}


\medskip  
		
\begin{center}
	{ \LARGE Supplementary material to the paper "Fully turbulent flows of viscoplastic fluids in a rectangular duct"}
\end{center}


\section{Verification of the LDA measurements}	\label{uncertainty}

In this section we assess the reliability of the LDA measurements. Similar verification procedures were performed in \citet{mohammadtabar2017} and \citet{tamano2018}, for example. As mentioned previously, the equipment calibration specify a 0.1\% measurement uncertainty. Due to small errors in alignment of the traverse and experimental setup, the wall normal velocity component $V$ deviates from zero by about 2\% of the mean streamwise velocity or less, for both water and all concentrations of Carbopol. Considering velocity bias errors in the LDA measurements, we used transit-time weighted averaging instead of simple time averaging if deviations greater than 2\% were encountered. A 5\% or slightly larger deviation between equally weighted data and trasit-time weighted data occured only near the wall, while the largest deviation for the root mean square of velocity fluctuations was near 2\% or lower. All Reynolds stress data for Carbopol was also corrected for velocity bias, with deviations from transit-time weighted averages being from 3 - 8\% approximately. Reynolds stresses of water experiments had minimal velocity bias, except for $\langle uv \rangle$ which was near 3\%. Skewness data had larger deviation when comparing time averaging and transit-time averaging, sometimes over 20\%. We thus performed corrections for the mean velocity and skewness data only. Formulations for transit-time weighted statistics can be found in \citet{owolabi2019}. Additionally, from the velocity profiles in the main paper, other sources of error affect the velocity measurements especially near the wall, such as gradients and the limitation of the 0.1 \textit{mm} diameter of the measurement volume.

Here we compare to pipe flow DNS measurements to our own LDA measurements in a pipe and the duct geometries with water. We use the pipe mainly as a verification for the LDA measurements. This is because LDA measurements of both streamwise and wall-normal components of velocity are possible with the duct, but we could not measure the wall-notmal velocity component with the pipe setup because of laser beam reflections due to the glass pipe curvature. The pipe main flow direction is the same as shown for the duct in the main paper. The circular test section is composed of five 1.5 $m$ borosilicate glass pipes with outer diameter (OD) of 60 $mm$ and inner diameter (ID) = $2R$ of 42 $mm$. The glass pipes are connected by aluminium blocks with the same ID of 42 $mm$ to avoid flow disturbance along the pipe. An acrylic box (or \textit{fish tank}) filled with glycerol is used to match the refractive index of the glass walls. For the experiments in the pipe we used a 300 \textit{mm} focal length lens which provides a 1 \textit{mm} length for the measurement volume. Therefore, the resolution near the wall is limited.

\begin{figure}
	\centering
	\includegraphics[width=\textwidth]{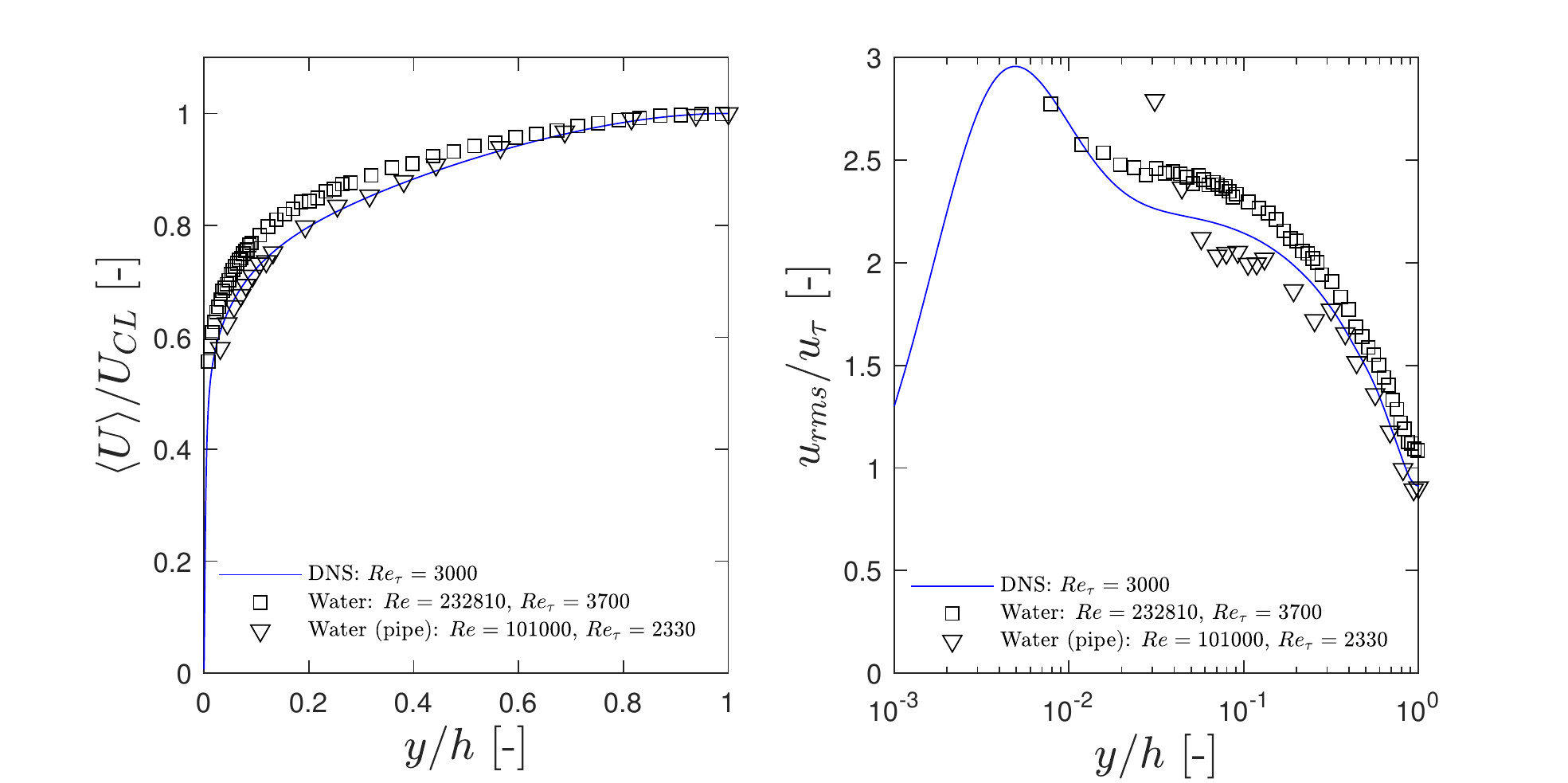}
	\caption{\fontsize{9}{9}\selectfont Velocity profiles normalized by the mean centreline velocity (a) streamwise turbulent intensities $u_{rms}$ normalized by the friction velocity (b) measurements of the turbulent flow of water in both pipe and duct geometries.} 
	\label{LDA_verification}
\end{figure}

We find the position of the wall in a manner similar to what was described in the experimental procedure in the main paper, but this time taking into account the glycerol and the acrylic of the fish tank in the calculation of the refraction of the beams. We only measure the streamwise velocity component. We traverse the probe backwards in the \textit{z}-direction, from the centreline of the pipe towards the wall, so the curvature of the pipe does not interfere with the laser beams. The verification by means of comparing $\langle U \rangle$ and $u_{rms}$ of measurements is shown in \hyperref[LDA_verification]{Figure~\ref*{LDA_verification}} (a) and (b) respectively, for both pipe and duct geometries. The time-averaged velocity profiles show a good agreement between all data shown, although there is a slight difference in $\langle U \rangle$ between the duct and pipe geometries from $y/h = 0.1$ to $0.5$, due to the effect of the 2:1 cross section of the duct. In \hyperref[LDA_verification]{Figure~\ref*{LDA_verification}} (b), the duct $u_{rms}$ measurements agrees better with the DNS pipe flow data than the duct experiments. The low spatial resolution of the pipe measurements is apparent in the $u_{rms}$ data, where reliable data is achieved for $y/h > 0.05$. Nevertheless, the good agreement between the experimental measurements in the different geometries confirm the reliability of the LDA experiments shown in the paper.

\begin{figure}
	\centering
	\includegraphics[width=\textwidth]{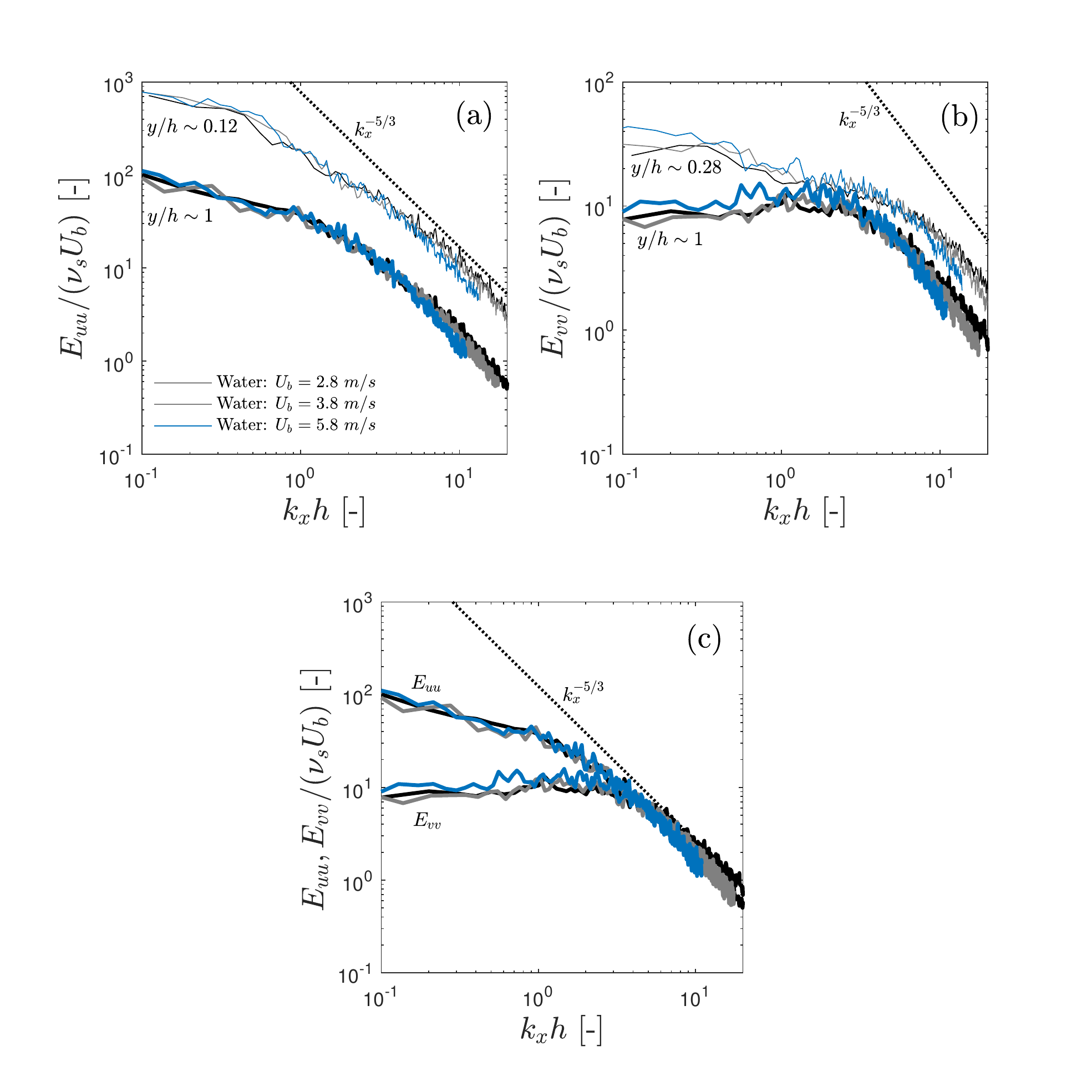}
	\caption{\fontsize{9}{9}\selectfont Power spectral densities of velocity fluctuations during the flow of water. (a) shows the PSDs of the streamwise velocity fluctuations at $y/h \sim 0.12$ (thin lines) and at the centre plane at $y/h \sim 1$ (thick lines). The $k_x^{-5/3}$ scaling is shown by dotted lines.  (b) shows the PSDs of the wall normal velocity fluctuations at $y/h \sim 0.28$ (thin lines) and at the centreline at $y/h \sim 1$ (thick lines). Image (c) shows the PSDs of streamwise ($E_{uu}$, thin lines) and wall normal velocity fluctuations ($E_{vv}$, thick lines) at the centreline $y/h \sim 1$.} 
	\label{water_spec1}
\end{figure}

As another benchmark, we show the PSDs of the $u$ fluctuations, given by $E_{uu}$, in \hyperref[water_spec1]{Figure~\ref*{water_spec1}} (a) and $v$ fluctuations  given by $E_{vv}$ in \hyperref[water_spec1]{Figure~\ref*{water_spec1}} (b) for water at three distinct Reynolds numbers at various $y/h$ positions specified in the figures. The dotted lines represent the wavenumber scaling for the inertial range of the turbulent energy cascade $k_x^{-5/3}$, where the energy transfer from large to small eddies depends only on the energy dissipation rate \citep{kolmogorov1941,pope2001}. Details on Taylor's hypothesis and data processing for estimating the PSDs are shown in the main paper. By means of scaling with the bulk velocity $U_b$ and the kinematic viscosity $\nu_s$, there is a very good collapse, both for $E_{uu}$ and $E_{vv}$. We observe that, for both positions shown in \hyperref[water_spec1]{Figure~\ref*{water_spec1}} (a), energy decay from small wavenumbers (representative of large eddies) to large wavenumbers (representative of small eddies) follow the $k_x^{-5/3}$ scaling quite closely for $k_x h > 5$. With \hyperref[water_spec1]{Figure~\ref*{water_spec1}}, we can establish that the experimental results are able to provide the energy spectra in the inertial range at reasonable wavenumbers. A similar conclusion is drawn from \hyperref[water_spec1]{Figure~\ref*{water_spec1}} (b), where the energy content from $v$ fluctuations also falls with the same $k_x^{-5/3}$ typically expected from Newtonian fluids at high $Re$. \hyperref[water_spec1]{Figure~\ref*{water_spec1}} (c) presents $E_{uu}$ and $E_{vv}$ at the centre of the duct. In this figure we observe a typical behaviour of isotropic turbulence (at least considering the measurement plane for our LDA results) in the centre of the duct, in which both $E_{uu}$ and $E_{vv}$ collapse in the high wavenumber range, also with approximately a $k_x^{-5/3}$ scaling \citep{cerbus2020}.

\begin{figure}
	\centering
	\includegraphics[width=\textwidth]{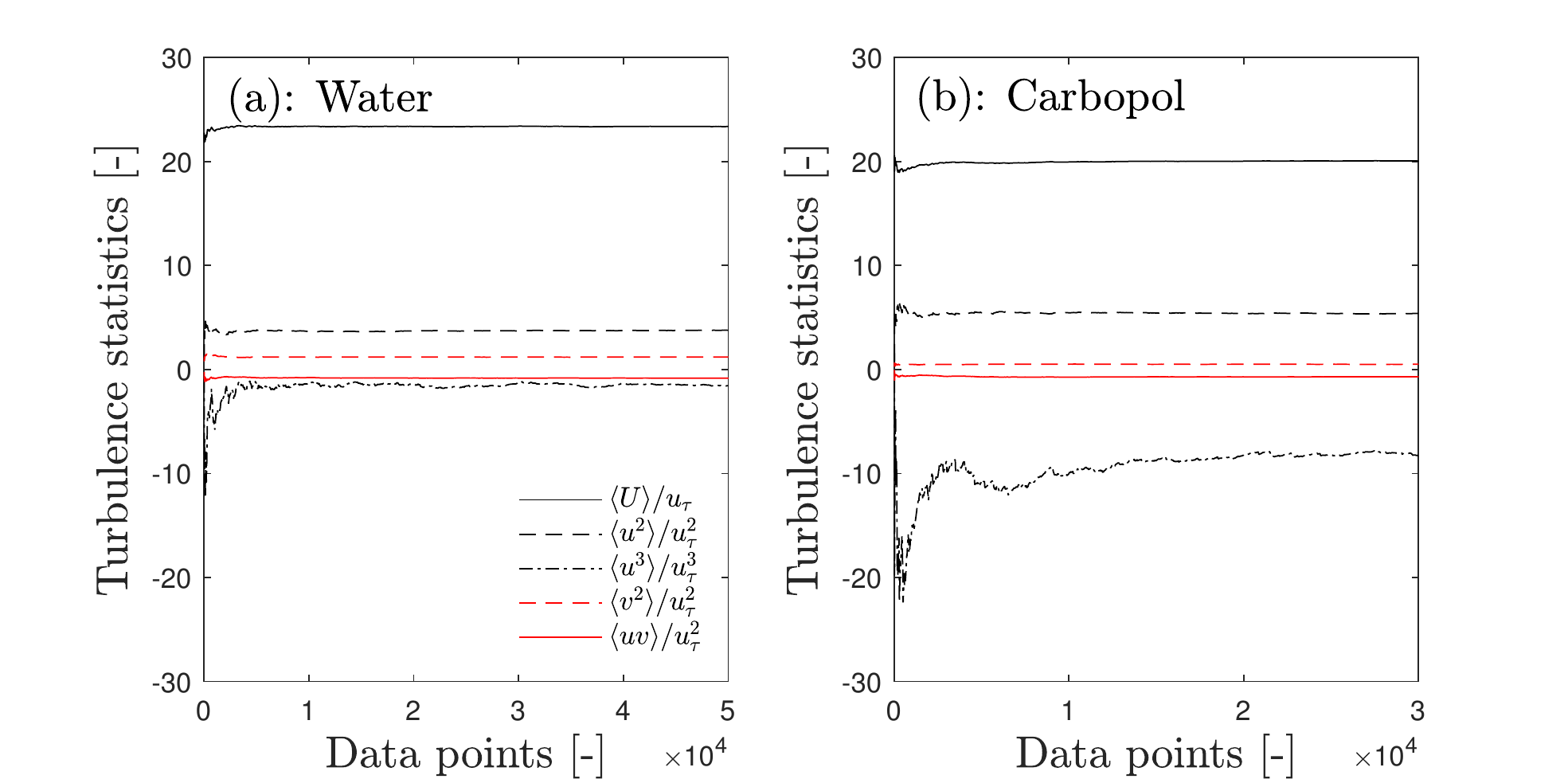}
	\caption{\fontsize{9}{9}\selectfont Statistical convergence of turbulent quantities for water (a) and 0.1\% Carbopol solution (b) at $y/h \sim 0.28$.} 
	\label{LDA_convergence}
\end{figure}

Next, we show the statistical convergence of measured quantities with the LDA in the duct geometry described in the main article, considering experiments with water in \hyperref[LDA_convergence]{Figure~\ref*{LDA_convergence}} (a) and the 0.1\% Carbopol solution in \hyperref[LDA_convergence]{Figure~\ref*{LDA_convergence}} (b). To account for both streamwise, wall normal and covariant statistics, we show data collected in coincidence mode, at $y/h \sim 0.28$. The objective is to verify the number of data points necessary for convergence of the statistical quantities. For this verification we consider the mean velocity $\langle U \rangle/u_{\tau}$, $\langle u^2\rangle/u_{\tau}^2$, $\langle v^2\rangle/u_{\tau}^2$, the Reynolds stresses $-\langle uv \rangle/u_{\tau}^2$ and $\langle u^3\rangle/u_{\tau}^3$. In the measured turbulence statistics of water in \hyperref[LDA_convergence]{Figure~\ref*{LDA_convergence}} (a), we observe a very good convergence of data at 10000 points, for all variables presented. Considering the 0.1\% Carbopol solution in \hyperref[LDA_convergence]{Figure~\ref*{LDA_convergence}} (b) the $\langle u^3\rangle/u_{\tau}^3$ data converges after 20000 data points, while the remaining turbulent quantities show good convergence much earlier, around 5000 points or less.

\begin{center}
	\bibliography{Results_carbopol}
	\linespread{1.00}
\end{center}